\documentclass{aa}  

\usepackage{graphicx}
\usepackage{caption}
\usepackage{subcaption}
\usepackage{comment}
\usepackage{txfonts}
\usepackage[bookmarks=false,colorlinks=true,linkcolor=black,citecolor=cyan,filecolor=black,urlcolor=cyan]{hyperref}
\usepackage{xcolor}

\begin{document}

   \title{The hot circumgalactic medium in the eROSITA All-Sky Survey}
   \subtitle{III. Star-forming and quiescent galaxies}
   
   \author{Yi Zhang \inst{1}\fnmsep\thanks{yizhang@mpe.mpg.de} \and
    Johan Comparat\inst{1} \and Gabriele Ponti\inst{2,1} \and Andrea Merloni\inst{1}
    \and Kirpal Nandra\inst{1}
    \and Frank Haberl\inst{1}
    \and Nhut Truong \inst{3,4}
    \and Annalisa Pillepich\inst{5}
    \and Paola Popesso\inst{6}
    \and Nicola Locatelli\inst{2}
    \and Xiaoyuan Zhang\inst{1}
    \and Jeremy Sanders\inst{1}
    \and Xueying Zheng\inst{1}
    \and Ang Liu\inst{1}
    \and Teng Liu\inst{7,8}
    \and Peter Predehl\inst{1}
    \and Mara Salvato\inst{1}
    \and Marcus Bruggen\inst{9} 
    \and Soumya Shreeram\inst{1}
    \and Michael C. H. Yeung\inst{1}
          }
          
   \institute{Max-Planck-Institut für extraterrestrische Physik (MPE), Gießenbachstraße 1, D-85748 Garching bei München, Germany
   \and
    INAF-Osservatorio Astronomico di Brera, Via E. Bianchi 46, I-23807 Merate (LC), Italy 
    \and
    NASA Goddard Space Flight Center, Greenbelt, MD 20771, USA 
    \and
     Center for Space Sciences and Technology, University of Maryland, 1000 Hilltop Circle, Baltimore, MD 21250, USA 
    \and
    Max-Planck-Institut für Astronomie, Königstuhl 17, 69117 Heidelberg, Germany 
    \and 
    European Southern Observatory, Karl Schwarzschildstrasse 2, D-85748 Garching bei München, Germany 
    \and
    Department of Astronomy, University of Science and Technology of China, Hefei 230026, China 
    \and 
    School of Astronomy and Space Science, University of Science and Technology of China, Hefei 230026, China
    \and Universität Hamburg, Hamburger Sternwarte, Gojenbergsweg 112, D-21029, Hamburg, Germany
             }
   \date{Received ; accepted }


  \abstract
  {}
   {The galaxy population shows a characteristic bimodal distribution based on the star formation activity and is sorted into star-forming or quiescent. These two subpopulations have a tendency to be located in different mass halos. The circumgalactic medium (CGM), as the gas repository for star formation, might contain the answer to the mystery of the formation of such bimodality. Here we consider the bimodality of the galaxy population and study the difference between the properties of the hot CGM around star-forming and quiescent galaxies.
   }
   {We used the X-ray data from the first four SRG/eROSITA all-sky surveys (eRASS:4). We selected central star-forming and quiescent galaxies from the Sloan Digital Sky Survey DR7 with stellar mass $10.0<\log(M_*/M_\odot)<11.5$ or halo mass $11.5<\log(M_{\rm 200m}/M_\odot)<14.0$ within spectroscopic redshift $z_{\rm spec}<0.2$ and build approximately volume-limited galaxy samples.
   We stacked the X-ray emission around star-forming and quiescent galaxies respectively. We masked detected point sources and carefully modeled the X-ray emission from unresolved active galaxy nuclei (AGN) and X-ray binaries (XRB) to detect the X-ray emission from the hot CGM. We measured the X-ray surface brightness ($S_{\rm X, CGM}$) profiles and integrated the X-ray emission from hot CGM within $R_{\rm 500c}$ ($L_{\rm X, CGM}$) to provide the scaling relations between $L_{\rm X, CGM}$ and galaxies' stellar or halo mass. 
   }
   {We detect extended X-ray emission from the hot CGM around star-forming galaxies with $\log(M_*/M_\odot)>11.0$ and quiescent galaxies with $\log(M_*/M_\odot)>10.5$, extending out to $R_{\rm 500c}$. The $S_{\rm X, CGM}$ profile of quiescent galaxies follows a $\beta$ model with $\beta \approx 0.4$, where $\beta$ quantifies the slope of the profile. $L_{\rm X, CGM}$ of star-forming galaxies with median stellar masses $\log(M_{\rm *,med}/M_\odot) = 10.7, 11.1, 11.3$ are approximately $0.8\,, 2.3\,, 4.0 \times 10^{40}\,\rm erg/s$, while for quiescent galaxies with $\log(M_{\rm *,med}/M_\odot) = 10.8, 11.1, 11.4$, they are $1.1\,, 6.2\,, 30 \times 10^{40}\,\rm erg/s$. Notably, quiescent galaxies with $\log(M_{\rm *,med}/M_\odot) > 11.0$ exhibit brighter hot CGM than their star-forming counterparts. In halo mass bins, we detect similar X-ray emission around star-forming and quiescent galaxies with $\log(M_{\rm 200m}/M_\odot) > 12.5$, suggesting that galaxies in the same mass dark matter halos host equally bright hot CGM. 
   We emphasize the observed $L_{\rm X, CGM} - M_{\rm 500c}$ relations of star-forming and quiescent galaxies are sensitive to the stellar-to-halo mass relation (SHMR). A comparison with cosmological hydrodynamical simulations (EAGLE, TNG100, and SIMBA) reveals varying degrees of agreement, contingent on the simulation and the specific stellar or halo mass ranges considered. 
   }
   {Either selected in stellar mass or halo mass, the star-forming galaxies do not host brighter stacked X-ray emission from the hot CGM than their quiescent counterparts at the same mass range. The result provides useful constraints on the extent of feedback's impacts as a mechanism for quenching star formation as implemented in current cosmological simulations. }


   \maketitle

\section{Introduction}
 
A puzzling observation that remains unexplained is the bimodality of the galaxy population.
In the color-magnitude diagram, early-type galaxies populate a narrow `red sequence’ that is separated by a shallow valley from a `blue cloud’ populated by late-type \citep[e.g.][]{Strateva2001,HoggBlantonEisenstein_2003ApJ...585L...5H,Baldry2004a,Baldry2006}. The red sequence galaxies are mostly elliptical (lenticular) and quiescent, while blue cloud galaxies actively form stars and appear predominantly disk-like. The emergence of quiescent galaxies within the framework of hierarchical galaxy formation has been a subject of interest. \citep[e.g.][]{FaberWillmerWolf_2007ApJ...665..265F, LaceyBaughFrenk_2016MNRAS.462.3854L}. 
Semi-analytic and hydrodynamic simulations have provided insights into the feedback processes can lead to galaxy quenching \citep{Croton2006,Somerville2008,Guo2011, Vogelsberger2014,Henriques2015,Schaye2015,Pillepich2018,Ma2022}. 
These simulations suggest that feedback from active galaxy nuclei (AGN) can play a dominant role in massive galaxies, where the energy released by the central black hole heats or expels the surrounding gas, effectively shutting down star formation. In lower-mass galaxies, stellar feedback from supernovae and stellar winds might be more significant. However, accurately modeling these feedback processes and their interplay remains challenging, necessitating new observational data to refine and constrain the models \citep{WechslerTinker_2018ARA&A..56..435W,Primack2024}. 

The bimodality is also evident in the stellar-to-halo-mass relation (SHMR), where quiescent galaxies at a given stellar mass reside in higher-mass dark matter halos than their star-forming counterparts \citep{Mandelbaum2006,More2011,Tinker2013,Velander2014,MandelbaumWangZu_2016MNRAS.457.3200M,TaylorCluverDuffy_2020MNRAS.499.2896T,Bilicki2021}. This bimodal SHMR suggests a strong connection between halo mass and the quenching of galaxies, highlighting how the environment and halo properties influence galaxy evolution. The higher halo masses associated with quiescent galaxies suggest that denser environments and deeper gravitational potentials might facilitate more efficient quenching mechanisms, such as stronger AGN feedback or more effective shock heating of the halo gas \citep{ZuMandelbaum_2016MNRAS.457.4360Z,BehrooziWechslerHearin_2019MNRAS.488.3143B,Moster2020,Cui2021,Shi2024}.

Despite the various observations and proposed quenching recipes  \citep{PuchweinSpringel_2013MNRAS.428.2966P, DuboisPichonWelker_2014MNRAS.444.1453D, HirschmannDolagSaro_2014MNRAS.442.2304H, Schaye2015, KhandaiDiMatteoCroft_2015MNRAS.450.1349K, DaveThompsonHopkins_2016MNRAS.462.3265D,NaabOstriker_2017ARA&A..55...59N, TumlinsonPeeplesWerk_2017ARAA..55..389T,Trussler2020,Chen2020,Guo2021,ZhangPeng2021,Hasan2023,Tchernyshyov2023}, the role and significance of each quenching mechanism remains elusive. 
The hot circumgalactic medium (CGM), closely associated with both feedback processes and halo mass, is a critical indicator for studying galaxy quenching \citep[i.e., as learnt from the intragroup and intracluster medium,][]{Lovisari2021, EckertGaspariGastaldello_2021Univ....7..142E}. The CGM consists of the hot, diffuse gas that surrounds galaxies and extends out to the virial radius (or beyond). The CGM is crucial in exchanging material between galaxies and their environments, influencing accretion and feedback processes. The properties of the hot CGM provide vital clues about the history of gas inflows and outflows and the effectiveness of various feedback mechanisms. A comprehensive understanding of the CGM's properties (X-ray luminosity, temperature, metallicity) and their relation to galaxy characteristics (specific star formation rate, stellar mass) provides essential constraints on quenching models \citep[see reviews by][]{NaabOstriker_2017ARA&A..55...59N, TumlinsonPeeplesWerk_2017ARAA..55..389T, EckertGaspariGastaldello_2021Univ....7..142E}.
Specifically, the scaling relation between stellar mass (or halo mass) and X-ray luminosity for star-forming and quiescent galaxies can provide significant insights into quenching mechanisms
\citep{OppenheimerBogdanCrain_2020ApJ...893L..24O,Truong2020, Truong2021b,Truong2021a}. 
Understanding where the SHMR bimodality occurs, particularly in galaxies with stellar masses around $\sim10^{10-11}M_\odot$, hosted by $M_{\rm vir}\sim10^{11.5-13}$M$_{\odot}$ halos is of great interest \citep{Moster2010, MosterNaabWhite_2013MNRAS.428.3121M, MosterNaabWhite_2018MNRAS.477.1822M, BehrooziConroyWechsler_2010ApJ...717..379B, BehrooziWechslerConroy_2013ApJ...770...57B, BehrooziWechslerHearin_2019MNRAS.488.3143B}.

However, studying the hot CGM around $\sim10^{10-11}M_\odot$ galaxies is challenging due to its diffuse and faint nature.
Pointed observations using {\it Chandra} and {\it XMM-Newton} have measured the hot gas properties of nearby ($<50$ Mpc) early-type and late-type galaxies, providing constraints on AGN and stellar feedback processes \citep{Ewan2001, Ewan2003, Strickland2004,Tullmann2006,David2006,Wang2010,Li2013a,Li2013b,Sarzi2013,Kim2013,Kim2015,Choi2015, Wang2016, Goulding2016,Forbes2017,Babyk2018,Kim2019, Hou2024}. 
With the PV/eFEDS observations of the extended ROentgen Survey with an Imaging Telescope Array (eROSITA) on board the Spektrum Roentgen Gamma (SRG) orbital observatory, \citet{Comparat2022} stacked $\sim$16,000, and \citet{Chada2022} stacked $\sim$1,600 star-forming and quiescent galaxies respectively, to study the X-ray emission around galaxies. Both works detect extended X-ray emission around quiescent and star-forming galaxies but suffer from satellite galaxy contamination and bias from the small sky area. 
This impeded drawing a clear conclusion on the implications of their measurements. 

This analysis is based on the state-of-the-art stacking results obtained in \citet[][hereafter Paper I]{Zhang2024profile}. 
Paper I applies the stacking method to the X-ray data on four eROSITA all-sky surveys (eRASS:4) \citep{Merloni2012,Predehl2021,SunyaevArefievBabyshkin_2021A&A...656A.132S,Merloni2024} around the approximately volume-limited central galaxy samples to measure their X-ray surface brightness profiles. 
Here,we split the central galaxy samples into star-forming (SF) and quiescent (QU) samples to investigate the differences in their X-ray profile and total X-ray luminosity. 

The eROSITA X-ray data reduction and stacking method are detailed in Paper I and \citet{Zhang2024relation}. This paper is organized as follows. 
The description of the sample compilation and a brief summary of the methods are given in Sect.~\ref{Sec_method}. 
The mean X-ray surface brightness profiles of the hot CGM and the scaling relations between integrated X-ray luminosity and stellar mass or halo mass for star-forming and quiescent galaxies are presented in Sect.~\ref{Sec_result}. 
We discuss the results in Sect.~\ref{Sec_discuss}. 
The conclusion is summarized in Sect.~\ref{Sec_summary}. 
Throughout, we use \citet{Planck2020} cosmological parameters: $H_0 = 67.74\ \rm km/s/Mpc$ and $\Omega_{\rm m} = 0.3089$. The $\log$ in this work designates $\log_{10}$. The stellar mass and halo mass in this work are in units of solar mass ($M_\odot$).


\section{Samples and stacking methods}\label{Sec_method}

\subsection{Galaxy samples} 

\begin{table}[h]
\begin{center}
\caption{Galaxy samples used in this work. }

\begin{tabular}{ccccccccccc}
\hline \hline 
\multicolumn{2}{c|}{$\rm log_{10}(M_*/M_\odot)$}&\multicolumn{2}{|c|}{redshift}&\multicolumn{2}{|c}{$N_{\rm gal}$}\\
min&max&min& max&CEN$_{\rm SF}$&CEN$_{\rm QU}$\\
\hline
10.0&10.5&0.01&0.06&5226&2516 \\
10.5&11.0&0.02&0.10&13668&16485 \\
11.0&11.25&0.02&0.15&8070&17466 \\
11.25&11.5&0.03&0.19&4279&15663 \\
\hline
\\
\end{tabular}

\begin{tabular}{cccccccccccccc} 
\hline \hline 
\multicolumn{2}{c|}{$\rm log_{10}(M_{\rm 200m}/M_\odot)$}&\multicolumn{2}{|c|}{redshift}&\multicolumn{2}{|c}{$N_{\rm gal}$}\\

min&max&min& max&CEN$_{\rm halo,SF}$&CEN$_{\rm halo,QU}$\\
\hline
11.5&12.0&0.01&0.08&20206&5162 \\
12.0&12.5&0.02&0.13&29791&11510 \\
12.5&13.0&0.02&0.16&12103&18637 \\
13.0&13.5&0.03&0.20&1788&18101 \\
13.5&14.0&0.03&0.20&-&8209 \\

\hline
\end{tabular}
\tablefoot{Top: Central star-forming (CEN$_{\rm SF}$) and quiescent (CEN$_{\rm QU}$) galaxy samples binned in stellar masses. From left to right, the columns are the stellar mass bin, redshift bin and number of star-forming (or quiescent) galaxies in the sample. Bottom: Central star-forming (CEN$_{\rm halo, SF}$) and quiescent galaxies (CEN$_{\rm halo, QU}$) binned in halo mass. From left to right, the columns are the halo mass bin, redshift bin, and the number of star-forming (or quiescent) galaxies in the sample. }
\label{Tab:gal:samples}
\end{center}
\end{table}

\begin{figure}[h]
    \centering
    \includegraphics[width=1.0\columnwidth]{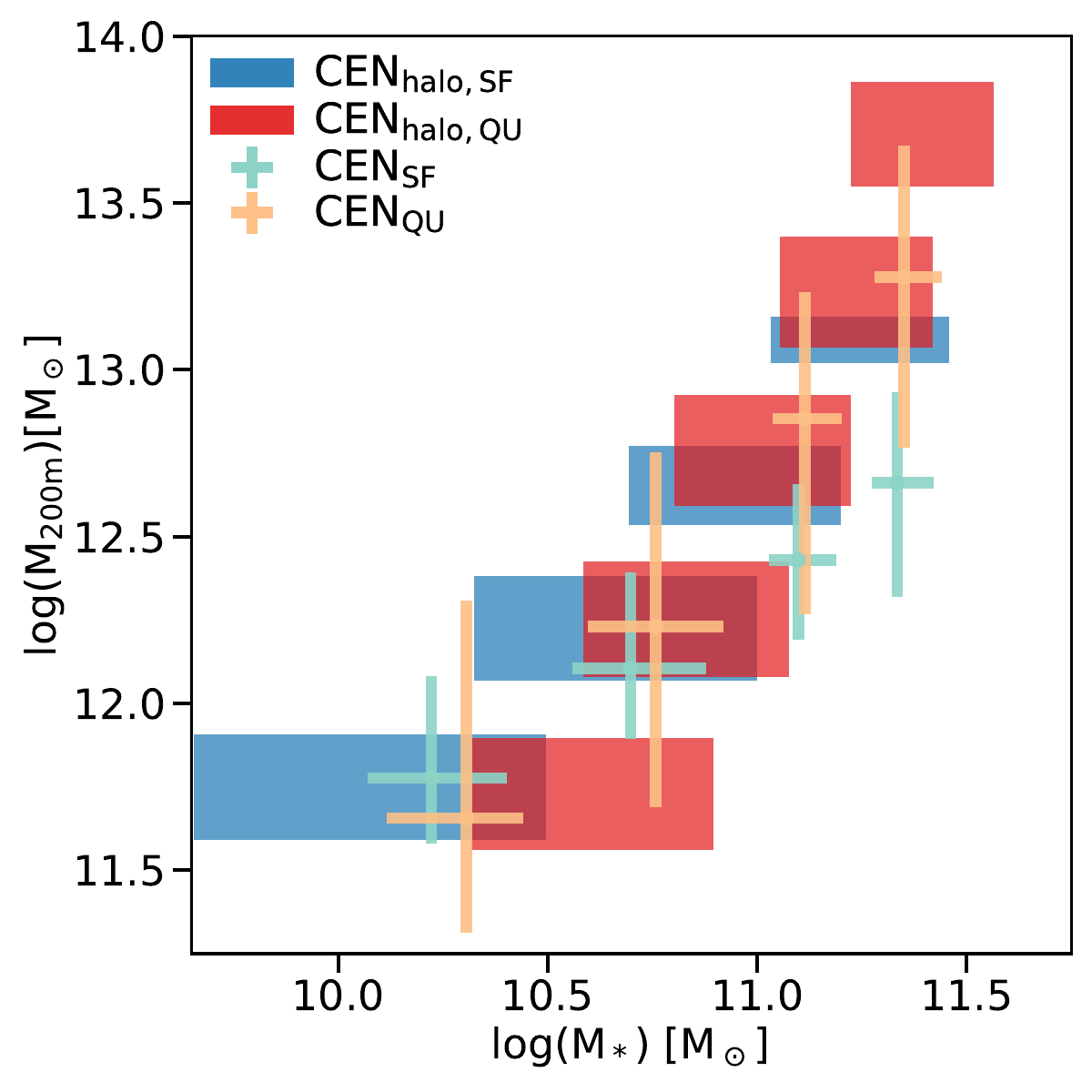}
    \caption{The median $M_{\rm 200m}$ and median $M_*$ of galaxies with the $16-84\%$ scatter for each mass bin in the CEN$_{\rm halo,SF}$ (blue band), CEN$_{\rm halo,QU}$ (red band), CEN$_{\rm SF}$ (green crosses), and CEN$_{\rm QU}$ (orange cross) samples.}
        \label{Fig_SHMR}
\end{figure}

We build the galaxy samples based on the SDSS DR7 spectroscopic galaxy catalog ($r_{\rm AB}<17.77$) \citep{Strauss2002}. We take the spectroscopic redshift ($z_{\rm spec}$) estimated by \citet{Blanton2005}, stellar mass estimated by \citet{Chen2012}, and star-formation rate (SFR) estimated by \citet{Brinchmann2004}. The star-forming or quiescent classification is defined based on the measurements of the 4000\AA \ break from \citet{Brinchmann2004}.
We study the hot CGM around the central galaxies only. The central galaxies are identified by the halo-based group finder of \citet{Tinker2021}, with a misclassification probability of about $2\%$ ($1\%$) for the central star-forming (quiescent) galaxies.
The group finder also assigns the halo mass ($M_{\rm 200m}$) to the central galaxies\footnote{The halo mass provided in the catalog \citet{Tinker2021} is $M_{\rm 200m}$, the mass within the radius where the mean interior density is 200 times the background universe density. To ease the comparison to literature, we convert the $M_{\rm 200m}$ to $M_{\rm 500c}$ with the mass-concentration relation model from \citet{Ishiyama2021}. $M_{\rm 500c}$ is the mass within $R_{\rm 500c}$, the radius where the density is 500 times the background universe critical density.} \citep{AlpaslanTinker_2020MNRAS.496.5463A,Tinker2021,Tinker2022}.
The group finder reproduces very closely summary statistics of galaxy clustering and galaxy-galaxy lensing, especially the bimodal SHMR of star-forming and quiescent galaxies \citep{ZehaviZhengWeinberg_2011ApJ...736...59Z,MandelbaumWangZu_2016MNRAS.457.3200M}.

We limit the maximum redshift of the galaxies, so the sample is approximately volume-limited and complete. 
We build four samples from SDSS DR7:
\begin{itemize}
    \item{CEN$_{\rm SF}$ sample}. This sample includes 32,190 star-forming galaxies (SF) in the stellar mass range $10.0<\log(M_*)<11.5$ and redshift $0.01<z_{\rm spec}<0.19$, binned as listed in Table~\ref{Tab:gal:samples}.
    \item{CEN$_{\rm QU}$ sample}. This sample includes 53,032 quiescent galaxies (QU) in the stellar mass range $10.0<\log(M_*)<11.5$ and redshift $0.01<z_{\rm spec}<0.19$.
    \item{CEN$_{\rm halo,SF}$ sample}. This sample includes 63,893 star-forming galaxies with $11.5<\log(M_{\rm 200m})<13.5$ and redshift $0.01<z_{\rm spec}<0.2$, binned as listed in Table~\ref{Tab:gal:samples}.
    \item{CEN$_{\rm halo,QU}$ sample}. This sample includes 61,619 quiescent galaxies with $11.5<\log(M_{\rm 200m})<14.0$ and redshift $0.01<z_{\rm spec}<0.2$.
\end{itemize}
The number of galaxies of each bin in the four samples is detailed in Table~\ref{Tab:gal:samples}. 

We present the median stellar mass ($M_{\rm *, med}$) and median halo mass ($M_{\rm 200m, med}$) with the $16-84\%$ scatter for each mass bin in the samples in Fig.~\ref{Fig_SHMR}. 
Due to the bimodal SHMR and its steep slope \citep[$M_{\rm 200m}\sim M_*^{1.6}$ for galaxies with $\log(M_*)\gtrapprox10.7$,][]{MandelbaumWangZu_2016MNRAS.457.3200M}, the star-forming and quiescent galaxies within the same $M*$ bin are hosted by halos with very different $M_{\rm 200m}$, as shown in Fig.~\ref{Fig_SHMR}. The bimodal feature is weaker at the low-mass end of SHMR \citep{MandelbaumWangZu_2016MNRAS.457.3200M}, together with the stronger upper-scattering from low-mass star-forming galaxies than low-mass quiescent galaxies, the quiescent galaxies have higher $M_*$ than their star-forming counterparts within the same halo mass bins, as shown in Fig.~\ref{Fig_SHMR} \citep[also see discussion in ][]{Moster2020}.
By having the samples selected in both $M_*$ and $M_{\rm 200m}$ bins, we can study if the different environment of star-forming and quiescent galaxies located would affect their hot CGM. 

\subsection{X-ray data reduction and stacking method}\label{Sec_model}

We used the eRASS:4 data (energy-calibrated event files, vignetting-corrected mean exposure map, source catalog) processed by the eROSITA Science Analysis Software System \citep[eSASS, version 020;][]{Brunner2022,Merloni2024}. We masked the sky area where $N_{\rm H}>10^{21} \rm cm^{-2}$, $E(B-V)<0.1$, or have overdense sources detected by eSASS (detail in Paper I). 

We calculated the physical distance ($R_{\rm kpc}$) between events and galaxies according to the redshifts of the galaxies and the angular separations. 
For all galaxies in the sample, we binned the events into histograms around the galaxies by $R_{\rm kpc}$, with a weight of 
\begin{equation}
    W_{\rm e}=E_{\rm obs} \frac{1}{t_{\rm exp}}\frac{f_{\rm nh} A_{\rm corr}\,4\pi D_L^2 }{ A_{\rm eff}}\ [\rm erg\,s^{-1}\,Mpc^2\,cm^{-2}], 
\end{equation}
where $E_{\rm obs}$ is the observed energy of the event, $t_{\rm exp}$ is the exposure time of the event, $f_{\rm nh}$ is the absorption correction, $A_{\rm corr}$ is the masking area correction, $D_{\rm L}$ is the luminosity distance to the galaxy, $A_{\rm eff}$ is the effective area of eROSITA.
We calculated the mean X-ray surface brightness ($S_{\rm X}$) as
\begin{equation}\label{Equ_Sx}
    S_{\rm X}= \sum_{r_0}^{r_1} \frac{ W_{\rm e} }{A_{\rm shell} N_{\rm g}}\ [\rm erg\,s^{-1}\,kpc^{-2}],
\end{equation} 
where $r_0$ and $r_1$ are the radial bins\footnote{The radial bin boundaries used are (0, 10, 30, 50, 75, 100, 200,  300, 400, 500, 600, 700, 800, 900, 1000, 1100, 1200, 1300, 1400, 1600, 1800, 2000, 2200, 2400, 2600, 2800, 3000) kpc.}, $A_{\rm shell}$ is the area of each radial bin and $N_{\rm g}$ is the number of galaxies stacked. 
We took the background ($S_{\rm X,bg}$) as the minimum value of $S_{\rm X}$ (calculated in each 100 kpc radial bin) beyond $R_{\rm 500c}$.
The justification of the background estimation is in the Appendix A of Paper I.
The mean clean X-ray surface brightness profile of the galaxy sample is calculated by $S_{\rm X, gal}=S_{\rm X}-S_{\rm X, bg}$.

The mean X-ray luminosity ($L_{\rm X}$) of the galaxy sample is calculated by integrating the X-ray emission within $R_{\rm 500c}$:
\begin{equation}
    L_{\rm X}= \Sigma_{0}^{R_{\rm 500c}} S_{\rm X, gal} \times A_{\rm shell}\ [\rm erg\,s^{-1}].
\end{equation}

Both $S_{\rm X, gal}$ and $L_{\rm X}$ are calculated in the $0.5-2$ keV energy bin (rest frame, soft X-ray). 
The uncertainties of $S_{\rm X, gal}$ and $L_{\rm X}$ are estimated by summing the squares of the Poisson error and the uncertainty estimated with Jackknife re-sampling (Paper I). 

We correct the X-ray contamination from the $2\%$ ($1\%$) misclassified centrals in the CEN and CEN$_{\rm halo}$ samples (paper I).
To detect the diffuse X-ray emission from the hot CGM, we masked detected X-ray point sources and subtracted the XRB and unresolved AGN contribution. 
We adopt the XRB model from \citet{Aird2017}, which relates the SFR and $M_*$ of the galaxy to its XRB luminosity ($L_{\rm XRB}$). 
To estimate the X-ray emission from unresolved AGN, we stacked the galaxies hosting optical AGN, identified by the BPT diagram, and calculated its maximal luminosity \citep[$L_{\rm AGN}$; Paper I,][]{BPT1981,Brinchmann2004}. As discussed in Appendix~\ref{Apd_agn}, the XRB and unresolved AGN in star-forming and quiescent galaxies are about $10^{39}-10^{40}\,\rm erg/s$, and star-forming galaxies host brighter XRB and AGN than quiescent galaxies.
We took the X-ray surface brightness profiles for the XRB and unresolved AGN in the stacked central galaxies in the shape of the point spread function (PSF) of eROSITA \citep[][Paper I]{Merloni2024}.
We used the mock catalog in Paper I to evaluate the satellite populations around the stacked central galaxies, with which we model the X-ray emission from XRB in the satellites. We neglected the contribution from X-ray AGN in satellite galaxies since they have a small occurrence  \citep[$<10$\%, see][]{Comparat2023}. 
We sum the models above and obtain the total X-ray luminosity from XRB and unresolved AGN ($L_{\rm X, AGN+XRB+SAT}$) and X-ray surface brightness profile ($S_{\rm X, AGN+XRB+SAT}$). 
The X-ray luminosity (surface brightness) of the hot CGM is therefore $L_{\rm X, CGM}=L_{\rm X}-L_{\rm X, AGN+XRB+SAT}$ ($S_{\rm X, CGM}=S_{\rm X, gal}-S_{\rm X, AGN+XRB+SAT}$). The uncertainty of the $L_{\rm X, CGM}$ is propagated from the uncertainties of $L_{\rm X}$ and $L_{\rm X, AGN+XRB+SAT}$, similar for $S_{\rm X, CGM}$.

\section{Results}
\label{Sec_result}

\subsection{X-ray surface brightness profiles in $M_{\rm *}$ bins and $L_{\rm X,CGM}$-$M_{\rm *}$ relation}\label{Sec_cen}

\begin{figure*}[h!tb]
    \centering
    \includegraphics[width=0.9\columnwidth]{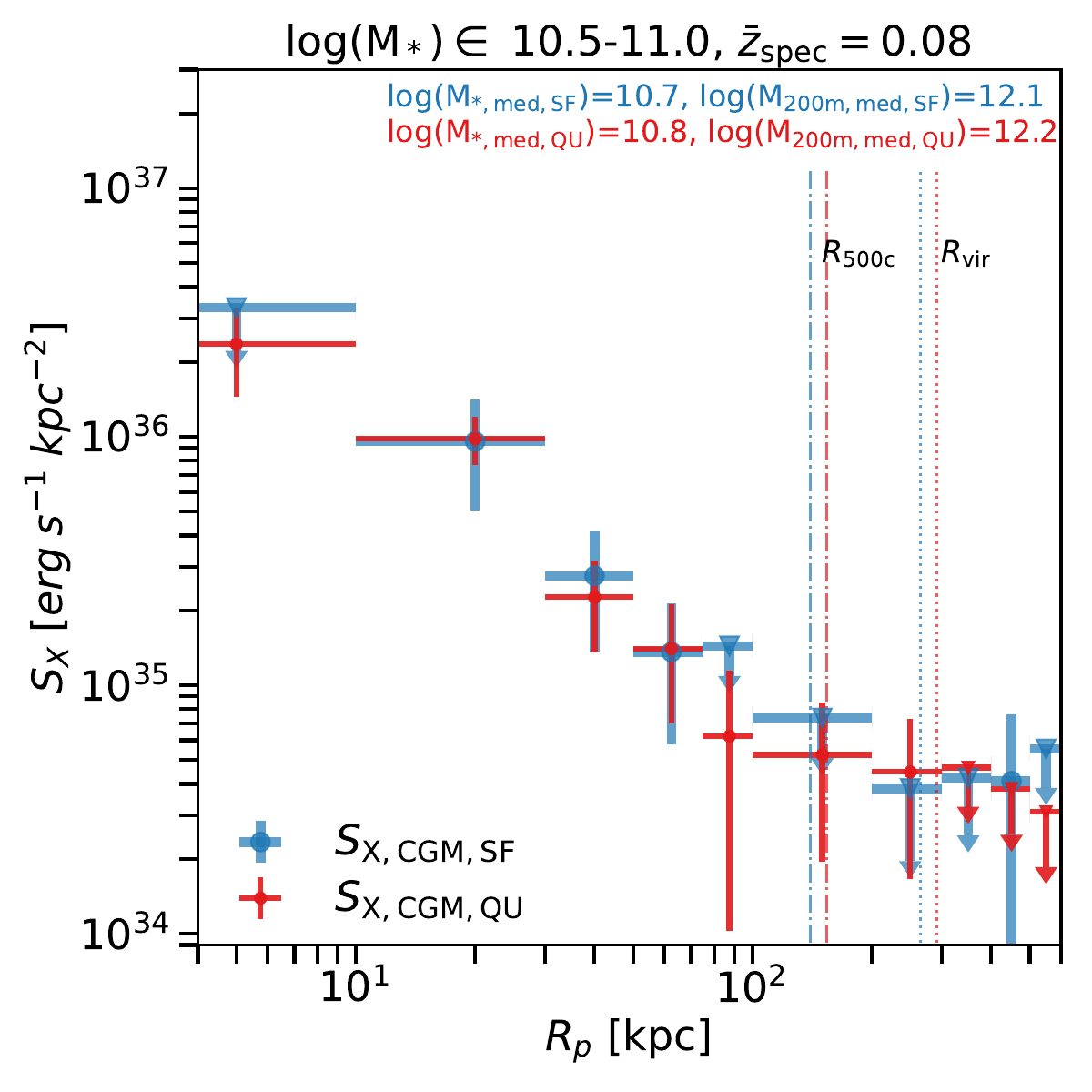}
    \includegraphics[width=0.9\columnwidth]{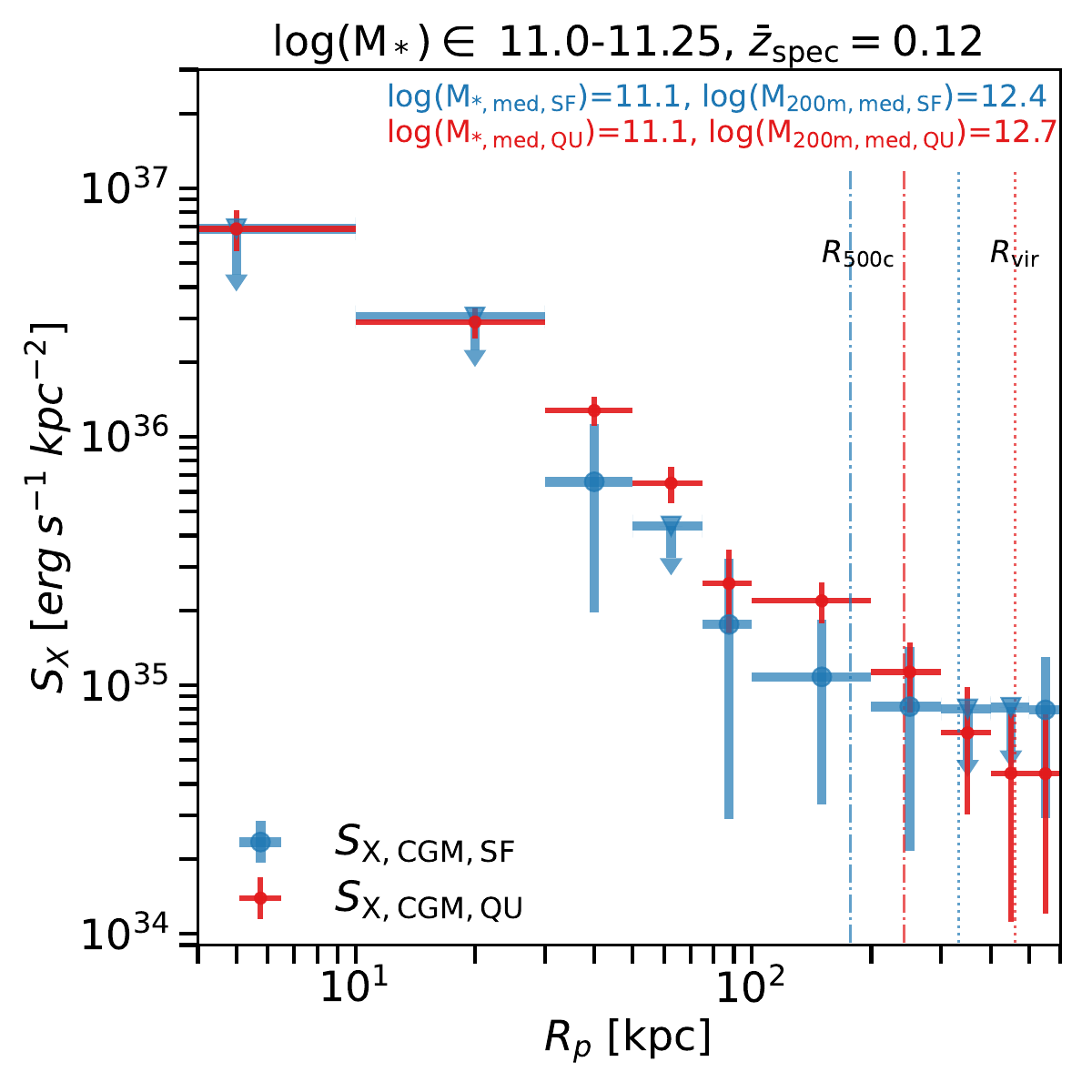} 
    \includegraphics[width=0.9\columnwidth]{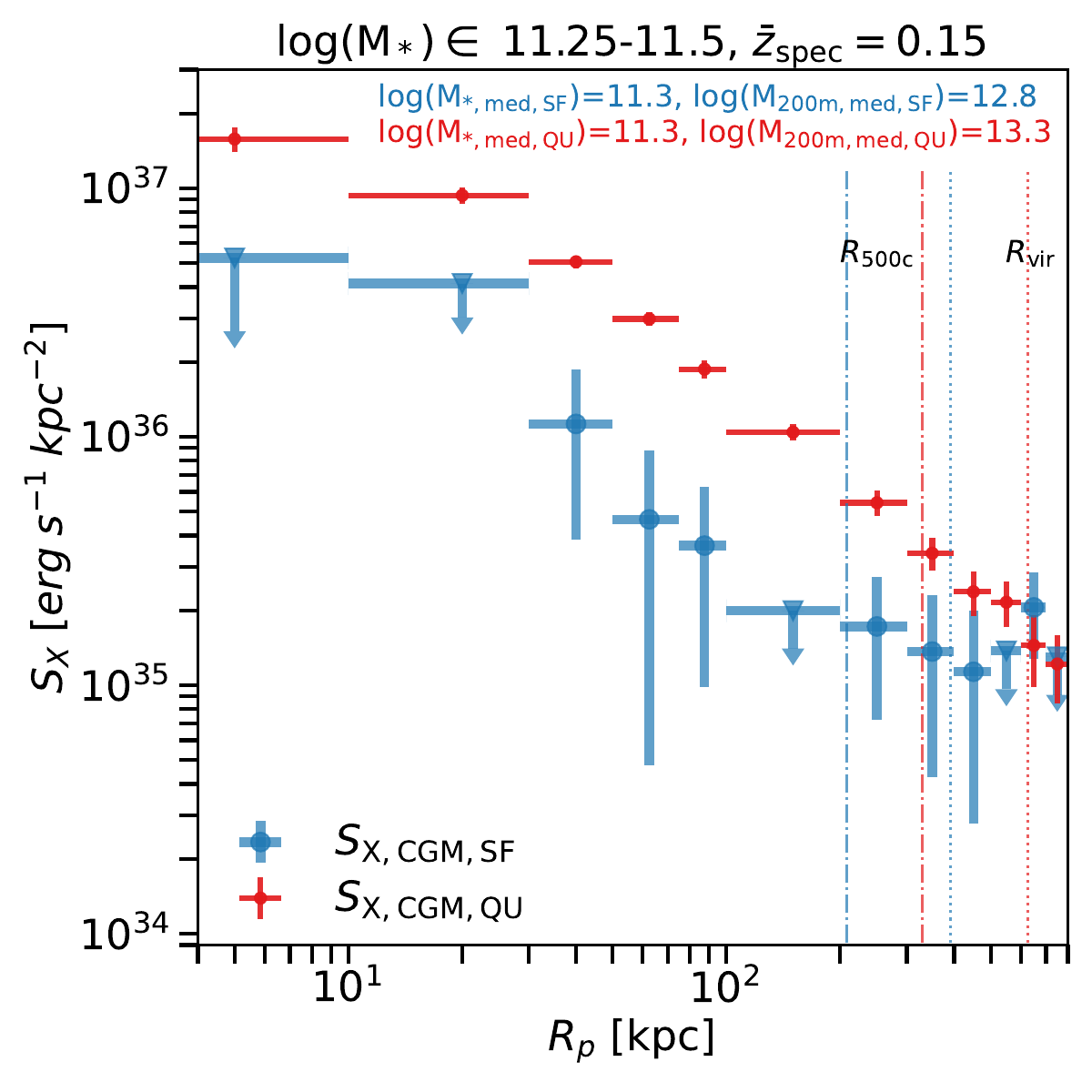}
    \includegraphics[width=0.9\columnwidth]{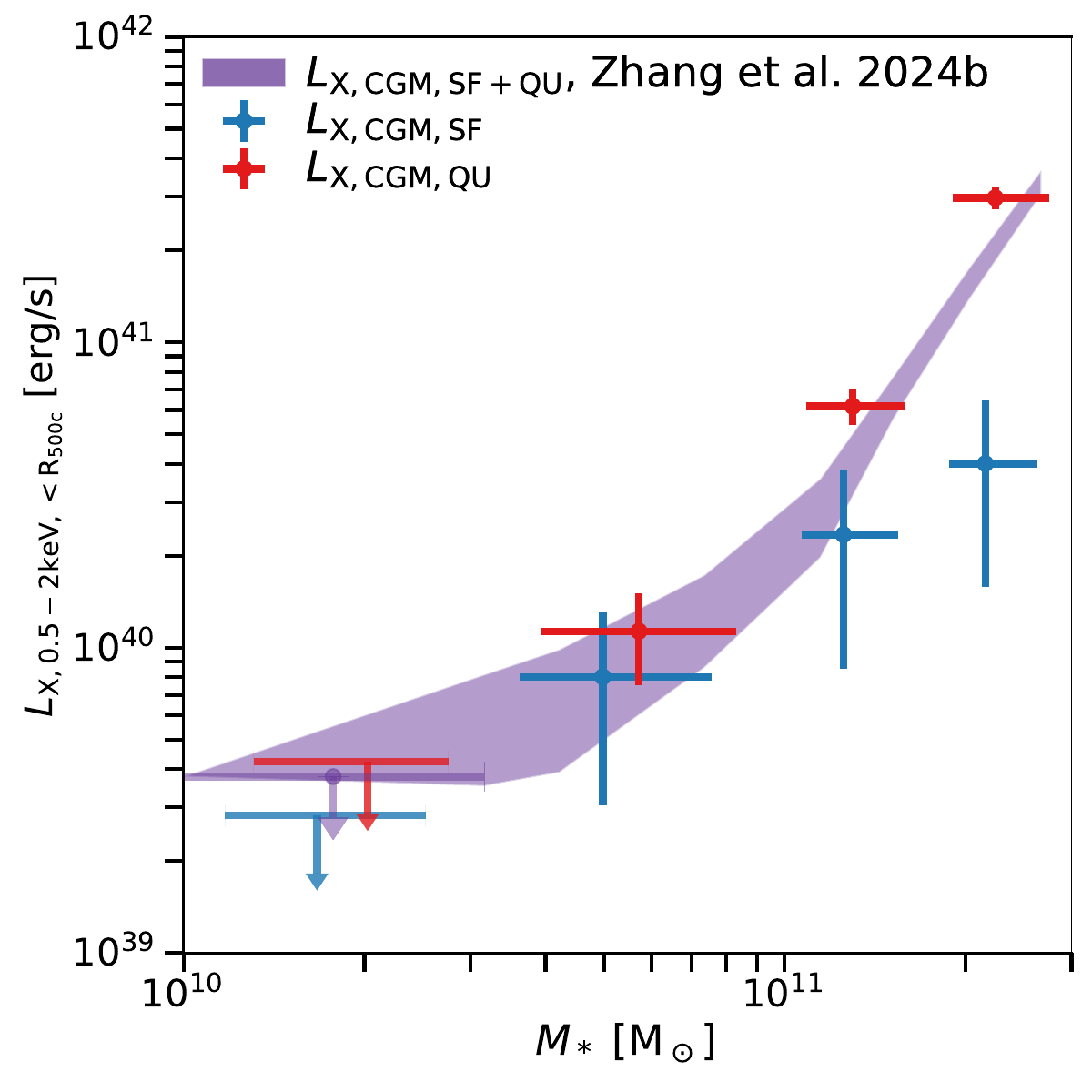}
\caption{X-ray surface brightness profiles of the hot CGM of the central star-forming ($S_{\rm X, CGM, SF}$, blue) and quiescent ($S_{\rm X, CGM, QU}$, red) galaxies with $M_{\rm *}\in 10.5-11.0$, $11.0-11.25$ and $11.25-11.5$ (top left, top right and bottem left). The vertical dash-dotted (dotted) lines denote the average $R_{\rm 500c}$ ($R_{\rm vir}$) of stacked central galaxies. The median redshift ($\bar{z}_{\rm spec}$), median $M_{\rm 200m}$ and median $M_*$ are denoted for star-forming and quiescent respectively. Hot CGM X-ray luminosity of central star-forming galaxies ($L_{\rm X,CGM, SF}$, blue) and quiescent galaxies ($L_{\rm X,CGM, QU}$,red) within $R_{\rm 500c}$ in $0.5-2$ keV as a function of the $M_{\rm *}$, compared to the $L_{\rm X,CGM}-M_*$ scaling relation from \citet{Zhang2024relation} (bottom right).}
        \label{Fig_profile_cen}
\end{figure*}

The $S_{\rm X,CGM}$ profiles for the central star-forming and quiescent galaxies with $\log(M_*)=10.5-11.5$ are plotted in Fig.~\ref{Fig_profile_cen}. 
The X-ray emission from the hot CGM increases with the stellar mass of the galaxy.
The extended X-ray emission from the hot CGM is detected out to $\approx R_{\rm 500c}$ around quiescent galaxies with $\log(M_*)>10.5$ and star-forming galaxies with $\log(M_*)>11.0$.
For star-forming and quiescent galaxies with $\log(M_*)=10.5-11.0$, we do not detect significant differences in their hot CGM X-ray emission. For galaxies with $\log(M_*)>11.0$, the quiescent galaxies have brighter and more extended hot CGM X-ray emission. 

We use a $\beta$ model \citep{Cavaliere1976} to describe the observed X-ray profiles of the hot CGM:

\begin{equation}
\label{eq:Sbeta}
    S_{\rm X,\,\beta}=S_{\rm X,\,0}\left[1+\left(\frac{r}{r_{\rm c}}\right)^2\right]^{-3\beta+\frac{1}{2}}\ [\rm erg\,s^{-1}\,kpc^{-2}],
\end{equation} 

\noindent where $S_{\rm X,0}$ is the X-ray surface brightness at the galaxy center, $r$ is the distance from the center, and $r_{c}$ is the core radius. We fit the PSF-convolved $\beta$ model to the observation using the Markov chain Monte Carlo (MCMC) method. 

The best-fit values and $1\sigma$ uncertainties of $\beta$, $S_{\rm X,0}$, and $r_{\rm c}$ are listed in Table~\ref{tab:betamodel}. 
For the quiescent galaxies, we obtain $\beta \approx 0.35-0.45$, $\log(S_{\rm X,0}) \approx 36.8-37.6$, poorly constrained $r_{\rm c}\approx 5$.
Point sources dominate the X-ray emission detected in star-forming galaxies; namely, the XRB and unresolved AGN contributes about 100\% X-ray emission in the lowest star-forming $\log(M_*)=10.0-10.5$ bin and about 50\% in $\log(M_*)=10.5-11.5$ bins, while XRB and AGN account for $<50\%$ for quiescent galaxies (see Table~\ref{table:LX:M}). Limited by the uncertainty of AGN and XRB models, we can not apply proper fits for the star-forming galaxies. 

The $L_{\rm X,CGM}-M_*$ scaling relation is plotted in the bottom right plot of Fig.~\ref{Fig_profile_cen}.
Above $\log(M_*)=11.0$, the quiescent galaxies host brighter hot CGM emission than star-forming galaxies, while no significant difference shown below $\log(M_*)=11.0$. 
We compare to the $L_{\rm X, CGM, SF+QU}$ obtained in \citet{Zhang2024relation}, where star-forming and quiescent galaxies are averaged together. $L_{\rm X, CGM, SF+QU}$ approaches to the $L_{\rm X, CGM}$ of quiescent galaxies at higher $M_*$, this agrees with the observational fact that quiescent fraction increases with $M_*$ \citep[e.g.,][]{Driver2022}.

\begingroup
\renewcommand{\arraystretch}{1.5} 
\begin{table*}[h!tb]
\centering
\caption{Best-fit parameters of $\beta$ model for the X-ray surface brightness profiles of CGM for the central star-forming and quiescent galaxies. }
\label{tab:betamodel}
\begin{tabular}{p{20mm}ccccccccccc}
\hline \hline
$\log(M_*)$&\multicolumn{1}{|c}{10.5-11.0}&\multicolumn{1}{|c}{11.0-11.25}&\multicolumn{1}{|c}{11.25-11.5}\\
&QU&QU&QU\\
\hline
$\beta$ &$0.45_{-0.07}^{+0.10}$&$0.37_{-0.02}^{+0.03}$&$0.35_{-0.01}^{+0.01}$ \\
$\log(S_{\rm X,0})$ \newline $[\rm erg\,s^{-1}\,kpc^{-2}]$&$36.8_{-0.4}^{+1.0}$&$37.7_{-0.5}^{+1.5}$&$37.6_{-0.1}^{+0.3}$\\
$r_{\rm c} [\rm kpc]$&$5_{-4}^{+8}$&$1_{-1}^{+4}$&$4_{-2}^{+2}$\\
\hline
\\
\end{tabular}

\begin{tabular}{p{20mm}ccccccccccc}
\hline \hline
$\log(M_{\rm 200m})$&\multicolumn{1}{|c}{12.0-12.5}&\multicolumn{1}{|c}{12.5-13.0}&\multicolumn{2}{|c}{13.0-13.5}&\multicolumn{1}{|c}{13.5-14.0}\\
&QU&QU&SF&QU&QU\\
\hline
$\beta$&$0.5_{-0.09}^{+0.07}$ &$0.4_{-0.03}^{+0.07}$&$0.51_{-0.1}^{+0.07}$&$0.42_{-0.03}^{+0.03}$&$0.40_{-0.02}^{+0.02}$ \\
$\log(S_{\rm X,0})$ \newline $[\rm erg\,s^{-1}\,kpc^{-2}]$&$37.0_{-0.5}^{+1.4}$&$37.2_{-0.3}^{+1.5}$&$37.4_{-0.6}^{+1.2}$&$37.5_{-0.1}^{+0.1}$&$37.7_{-0.1}^{+0.1}$\\
$r_{\rm c} [\rm kpc]$&$6_{-5}^{+9}$&$4_{-3}^{+7}$&$18_{-14}^{+27}$&$11_{-4}^{+4}$&$14_{-3}^{+4}$\\
\hline
\\
\end{tabular}

\tablefoot{The uncertainty is of 1$\sigma$. We use the X-ray surface brightness profiles within $R_{\rm vir}$ for the fits.}
\end{table*}
\endgroup

\subsection{X-ray surface brightness profiles in halo mass bins and the $L_{\rm X,CGM}$-$M_{\rm 500c}$ scaling relation}\label{Sec_cenhalo}

\begin{figure*}[h!tb]
    \centering
    \includegraphics[width=0.9\columnwidth]{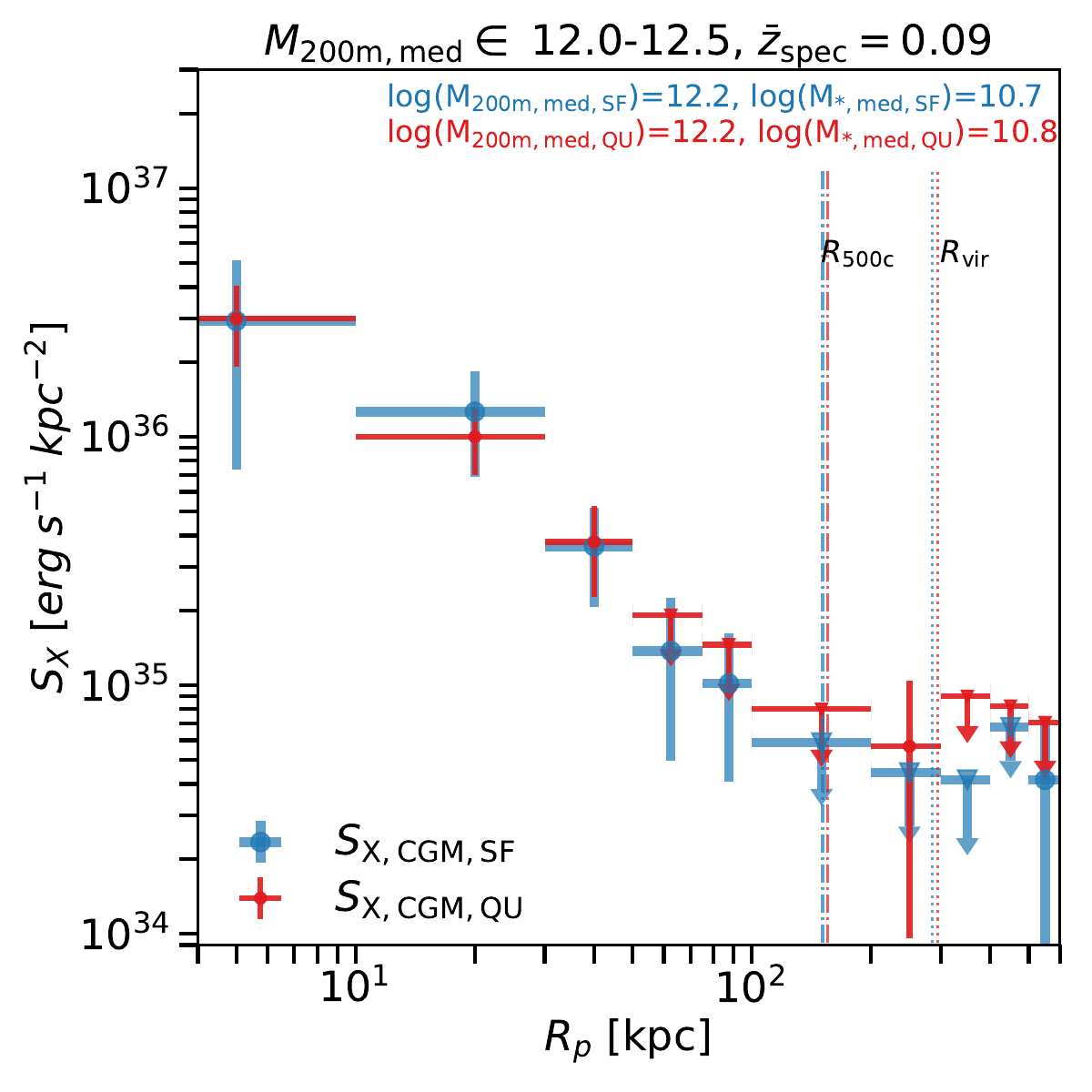}
    \includegraphics[width=0.9\columnwidth]{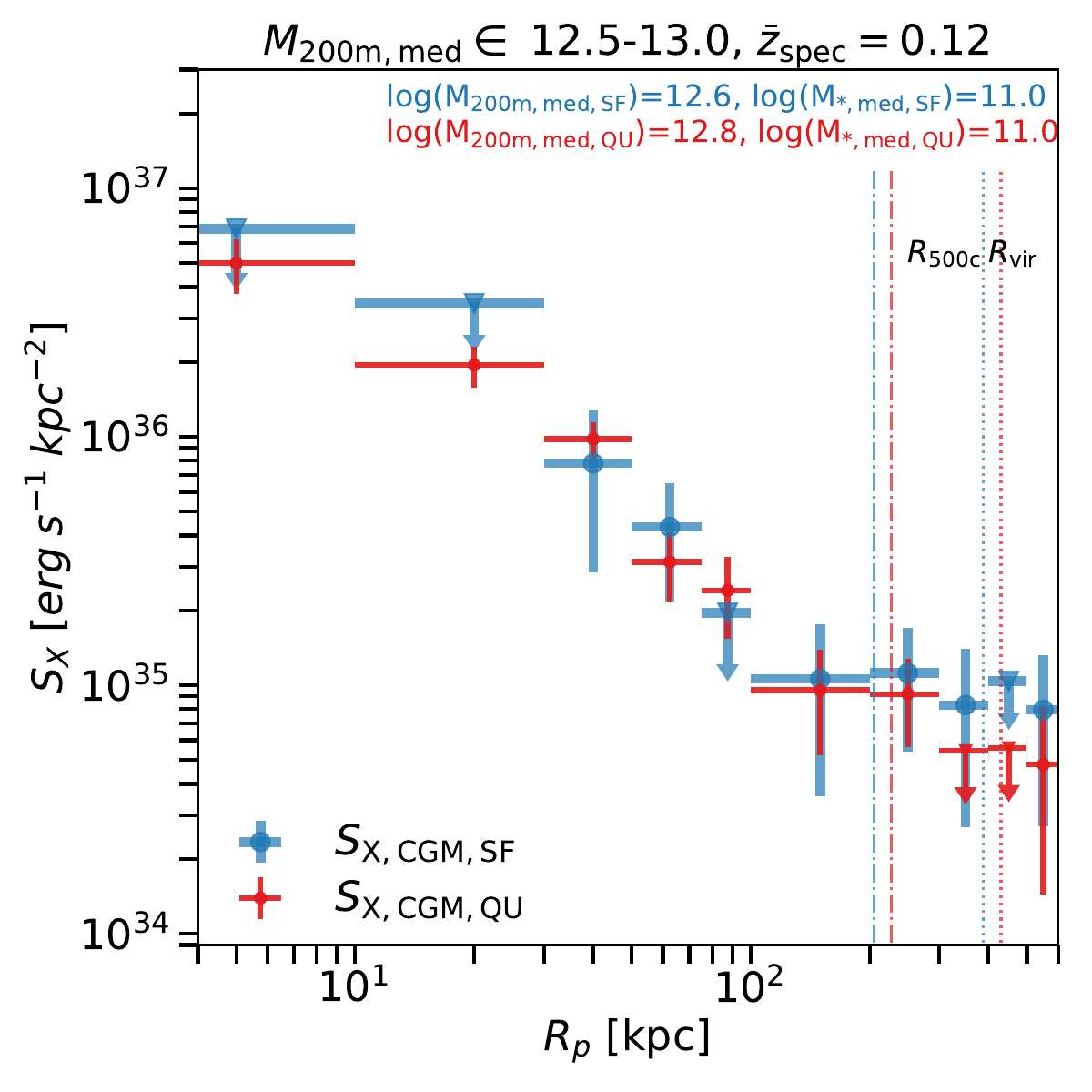}
    \includegraphics[width=0.9\columnwidth]{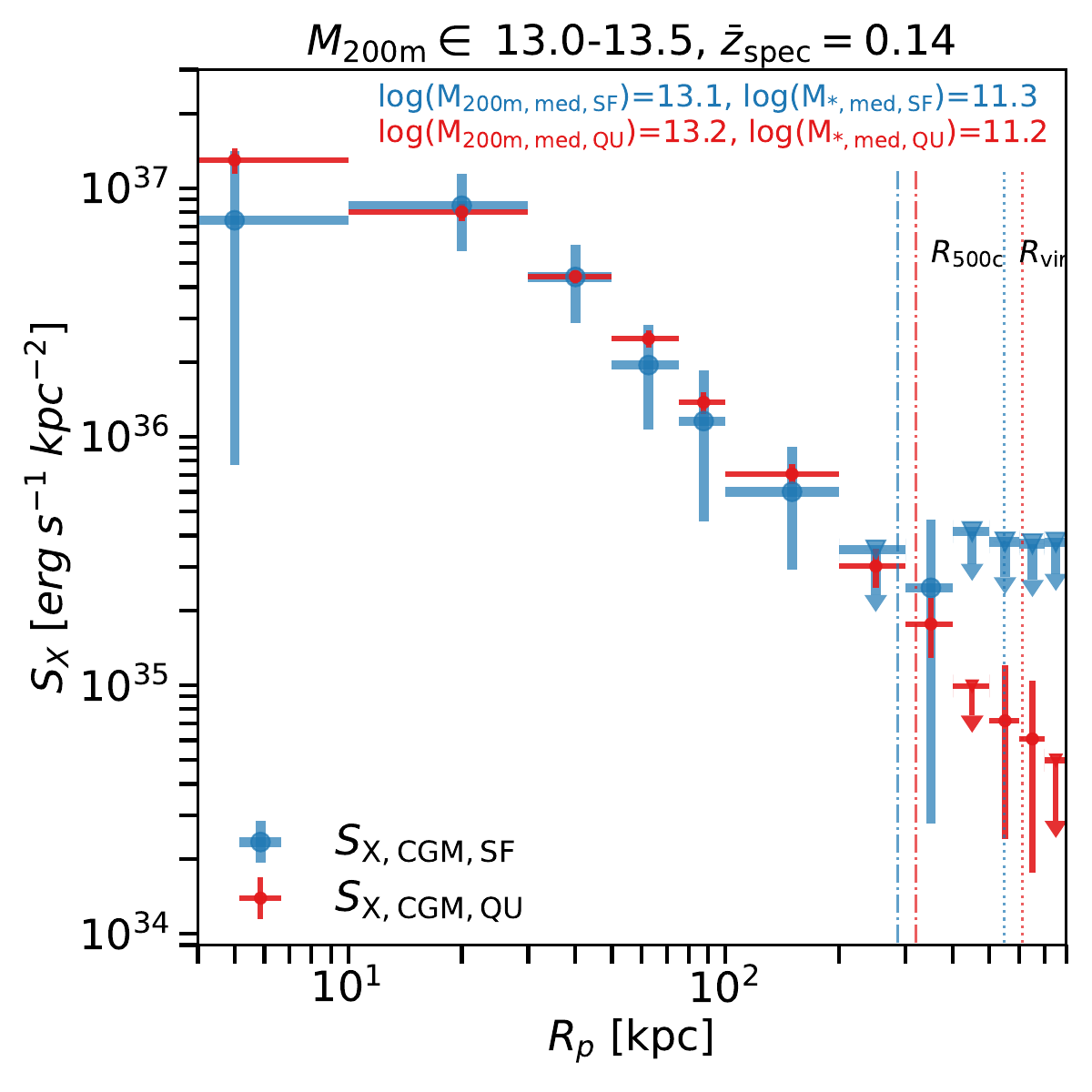}
    \includegraphics[width=0.9\columnwidth]{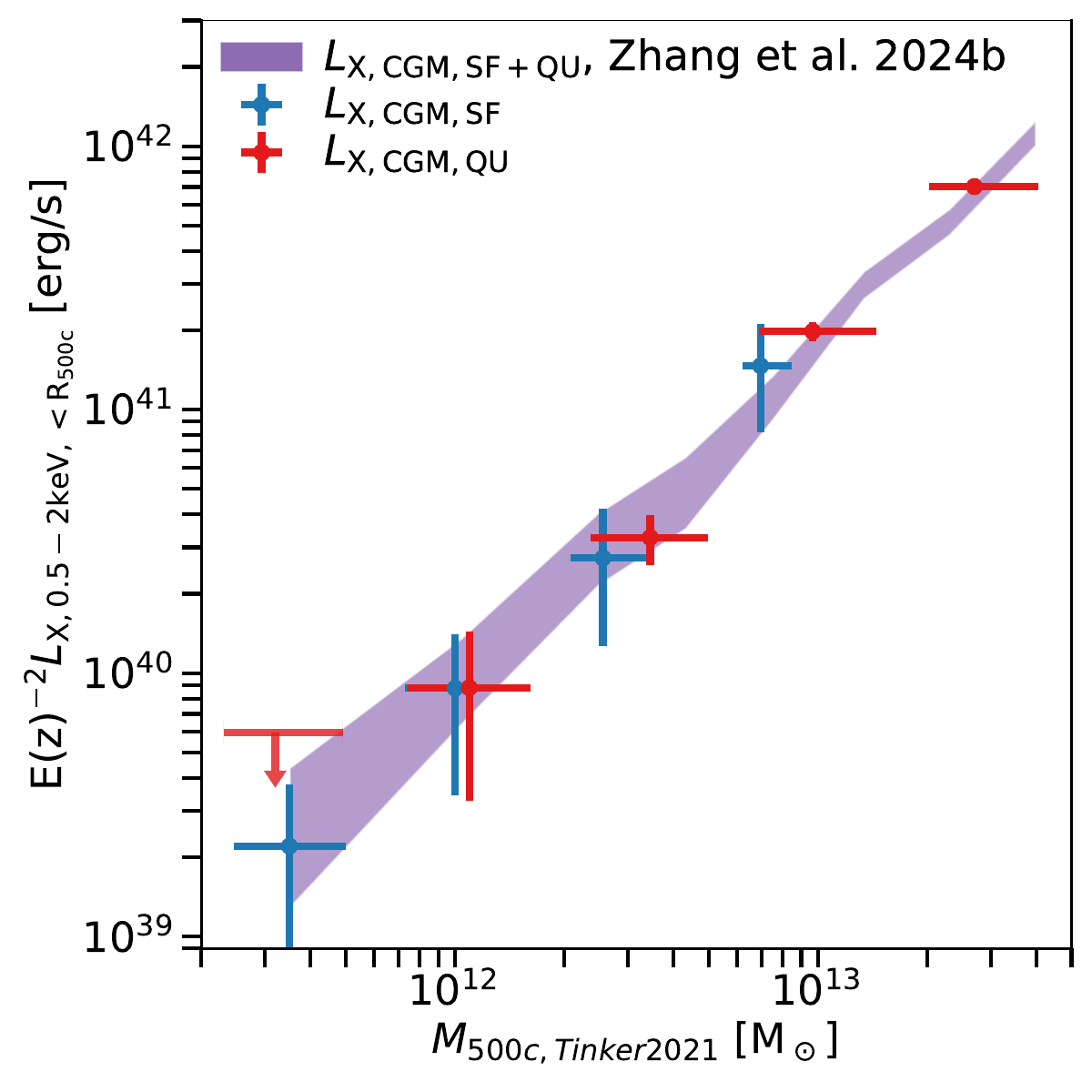}
\caption{X-ray surface brightness profiles of the hot CGM of the central star-forming ($S_{\rm X, CGM, SF}$, blue) and quiescent ($S_{\rm X, CGM, QU}$, red) galaxies with $M_{\rm 200m}\in 12.0-12.5$, $12.5-13.0$ and $13.0-13.5$ (top left, top right and bottem left). The vertical dash-dotted lines denote the average $R_{\rm 500c}$ and $R_{\rm vir}$ of stacked galaxies. The median redshift ($\bar{z}_{\rm spec}$), median $M_{\rm 200m}$ and median $M_*$ are denoted. Hot CGM X-ray luminosity of central star-forming galaxies ($L_{\rm X,CGM, SF}$, blue) and quiescent galaxies ($L_{\rm X,CGM, QU}$, red) within $R_{\rm 500c}$ in $0.5-2$ keV as a function of $M_{\rm 500c}$, compared to the $L_{\rm X,CGM}-M_{\rm 500c}$ scaling relation from \citet{Zhang2024relation} (bottom right). }
        \label{Fig_profile_cenhalo}
\end{figure*}

We present in the top panel and bottom left of Fig.~\ref{Fig_profile_cenhalo} the $S_{\rm X, CGM}$ profiles of star-forming and quiescent galaxies with halo mass ranges in $\log(M_{\rm 200m})=12.0-13.5$. 
We detect extended X-ray emission from the hot CGM out to $R_{\rm 500c}$ around star-forming and quiescent galaxies with $\log(M_{\rm 200m})>12.5$. The $S_{\rm X, CGM}$ of star-forming and quiescent are consistent within the uncertainties. This finding suggests that quiescent and star-forming central galaxies residing in similar-mass dark matter halos exhibit comparably bright hot CGM emission. 
We fit the $\beta$ model to the $S_{\rm X, CGM}$ and the parameters are listed in Table~\ref{tab:betamodel}. In general, we find $\beta\approx0.4-0.5$.

In the bottom right of Fig.~\ref{Fig_profile_cenhalo},  we present the $L_{\rm X,CGM}-M_{\rm 500c}$ scaling relations of central star-forming and quiescent galaxies. The similar $L_{\rm X,CGM}-M_{\rm 500c}$ scaling relations of central star-forming and quiescent galaxies follow approximately the $\log(L_{\rm X, CGM, SF+QU})= (1.32 \pm 0.05) \log(M_{\rm 500c}) + (24.1 \pm 0.7)$ obtained in \citet{Zhang2024relation}, where there is no split of star-forming and quiescent galaxies.

\begin{table*}[h!tb]
\centering
\caption{Rest-frame X-ray luminosity in the 0.5-2 keV band within $R_{\rm 500c}$ in erg s$^{-1}$ of different components for CEN and CEN$_{\rm halo}$ samples.}
\begin{tabular}{ccccccccccc}
\hline \hline
\multicolumn{3}{c}{$\log_{10}(M_*/M_\odot)$}&\multicolumn{4}{|c}{CEN$_{\rm SF}$}\\
min&max&med&$L_{\rm X,total}$&$L_{\rm X,mask}$&$L_{\rm X,CGM}$& $L_{\rm X,AGN+XRB+SAT}$ \\
&&&(no mask)&(with mask)&&(with mask)\\
\hline
10.0&10.5&10.2&$8.9\pm7.1\times10^{39}$&$2.5\pm2.9\times10^{39}$&-&$2.7\pm0.8\times10^{39}$ \\
10.5&11.0&10.7&$3.6\pm0.9\times10^{40}$&$1.7\pm0.4\times10^{40}$&$8.0\pm5.0\times10^{39}$&$9.2\pm2.9\times10^{39}$ \\
11.0&11.25&11.1&$1.3\pm0.4\times10^{41}$&$5.0\pm1.0\times10^{40}$&$2.3\pm1.5\times10^{40}$&$2.6\pm1.1\times10^{40}$ \\
11.25&11.5&11.3&$2.7\pm0.6\times10^{41}$&$7.8\pm1.8\times10^{40}$&$4.0\pm2.4\times10^{40}$&$3.8\pm1.7\times10^{40}$\\
       \hline
\multicolumn{2}{c}{}&\multicolumn{4}{c}{CEN$_{\rm QU}$}\\   
\hline
10.0&10.5&10.3&-&$2.9\pm2.7\times10^{39}$&$1.5\pm2.7\times10^{39}$&$1.4\pm0.6\times10^{39}$ \\
10.5&11.0&10.8&$2.6\pm0.9\times10^{40}$&$1.6\pm0.4\times10^{40}$&$1.1\pm0.4\times10^{40}$&$4.4\pm1.0\times10^{39}$ \\
11.0&11.25&11.1&$1.2\pm0.3\times10^{41}$&$7.3\pm0.8\times10^{40}$&$6.2\pm0.8\times10^{40}$&$1.1\pm0.3\times10^{40}$ \\
11.25&11.5&11.35&$4.4\pm0.6\times10^{41}$&$3.2\pm0.2\times10^{41}$&$3.0\pm0.2\times10^{41}$&$2.1\pm0.4\times10^{40}$\\
\hline
\\
\end{tabular}

\begin{tabular}{ccccccccccc}
\hline \hline
\multicolumn{3}{c}{$\log_{10}(M_{\rm 200m}/M_\odot)$}&\multicolumn{4}{|c}{CEN$_{\rm halo,SF}$}\\
min&max&med&$L_{\rm X,total}$&$L_{\rm X,mask}$&$L_{\rm X,CGM}$& $L_{\rm X,AGN+XRB+SAT}$ \\
&&&(no mask)&(with mask)&&(with mask)\\
\hline

11.5&12.0&11.75&$5.3\pm3.6\times10^{39}$&$4.5\pm1.6\times10^{39}$&$2.3\pm1.7\times10^{39}$&$2.2\pm0.5\times10^{39}$ \\
12.0&12.5&12.2&$4.1\pm0.9\times10^{40}$&$2.1\pm0.4\times10^{40}$&$9.6\pm5.8\times10^{39}$&$1.2\pm0.4\times10^{40}$ \\
12.5&13.0&12.6&$1.9\pm0.4\times10^{41}$&$5.9\pm1.1\times10^{40}$&$3.1\pm1.6\times10^{40}$&$2.8\pm1.3\times10^{40}$ \\
13.0&13.5&13.1&$6.5\pm2.5\times10^{41}$&$1.9\pm0.7\times10^{41}$&$1.7\pm0.7\times10^{41}$&$2.4\pm1.8\times10^{40}$\\
\hline
\multicolumn{2}{c}{}&\multicolumn{4}{c}{CEN$_{\rm halo,QU}$}\\
\hline
11.5&12.0&11.7&$2.7\pm2.1\times10^{40}$&$6.2\pm3.3\times10^{39}$&$2.7\pm3.6\times10^{39}$&$3.5\pm1.5\times10^{39}$ \\
12.0&12.5&12.25&$3.6\pm1.2\times10^{40}$&$1.4\pm0.6\times10^{40}$&$9.6\pm6.0\times10^{39}$&$4.9\pm1.5\times10^{39}$ \\
12.5&13.0&12.8&$6.7\pm1.9\times10^{40}$&$4.6\pm0.8\times10^{40}$&$3.7\pm0.8\times10^{40}$&$9.1\pm2.2\times10^{39}$ \\
13.0&13.5&13.2&$3.2\pm0.5\times10^{41}$&$2.4\pm0.2\times10^{41}$&$2.3\pm0.2\times10^{41}$&$1.7\pm0.3\times10^{40}$\\
13.5&14.0&13.7&$1.1\pm0.1\times10^{42}$&$8.6\pm0.6\times10^{41}$&$8.2\pm0.6\times10^{41}$&$3.7\pm0.7\times10^{40}$\\
\hline
\end{tabular}

\tablefoot{For the CEN$_{\rm SF}$ and CEN$_{\rm QU}$ samples (top), CEN$_{\rm halo,SF}$ and CEN$_{\rm halo,QU}$ samples (bottom), the X-ray luminosity of all X-ray emission ($L_{\rm X,total}$, without masking detected X-ray sources), the X-ray luminosity after excluding detected point sources ($L_{\rm X,mask}$, with masking X-ray point sources), the X-ray luminosity of hot CGM ($L_{\rm X,CGM}$), and modeled X-ray luminosity from unresolved AGN, XRB sources ($L_{\rm X,AGN+XRB+SAT}$). }
\label{table:LX:M}
\end{table*}


\section{Discussion}\label{Sec_discuss}
In Sect.~\ref{Sec_lxm}, we discuss how the uncertainty of halo mass assigned to galaxies affects the $L_{\rm X,CGM}-M_{\rm 500c}$ relation. 
In Sect.~\ref{Sec_sim}, we compare the observed $L_{\rm X,CGM}-M_{\rm 500c}$ and $L_{\rm X,CGM}-M_{\rm *}$ scaling relations to the hydrodynamical simulations.
In Sect.~\ref{Sec_sfr}, we split star-forming and quiescent by the specific SFR and find it does not change the conclusion. 
In Sect.~\ref{Sec_lit}, we compare our results to \citet{Comparat2022} and \citet{Chada2022} and explain why different conclusions were drawn in the two works.

\subsection{Effect of SHMR on $L_{\rm X,CGM}$-$M_{\rm 500c}$}\label{Sec_lxm}

In Sect.~\ref{Sec_cenhalo}, we find consistent $L_{\rm X,CGM}$-$M_{\rm 500c}$ scaling relations for star-forming and quiescent galaxies, which implies that, to the first order, halo mass is the determining factor for the CGM heating. The different feedback processes in star-forming and quiescent galaxies do not cause an observable difference in the hot CGM X-ray emission, which suggests that the regulation of feedback activity on CGM is tightly correlated with halo mass.

However, the halo masses of the galaxies in the SDSS catalog is not directly observable but is assigned by different group finder models \citep{Yang2007, Robotham2011, Tinker2022,Zhao2024}. The halo mass ($M_{\rm 500c}$ or $M_{\rm 200m}$) used in this work is provided by \citet{Tinker2021}, where the halo mass is assigned according to two parameters: total luminosity of the galaxy group and its concentration. The halo mass is calibrated to obtain consistent bimodal SHMR found in \citet{MandelbaumWangZu_2016MNRAS.457.3200M}. 

\begin{figure}[h!tb]
    \centering
    \includegraphics[width=0.9\columnwidth]{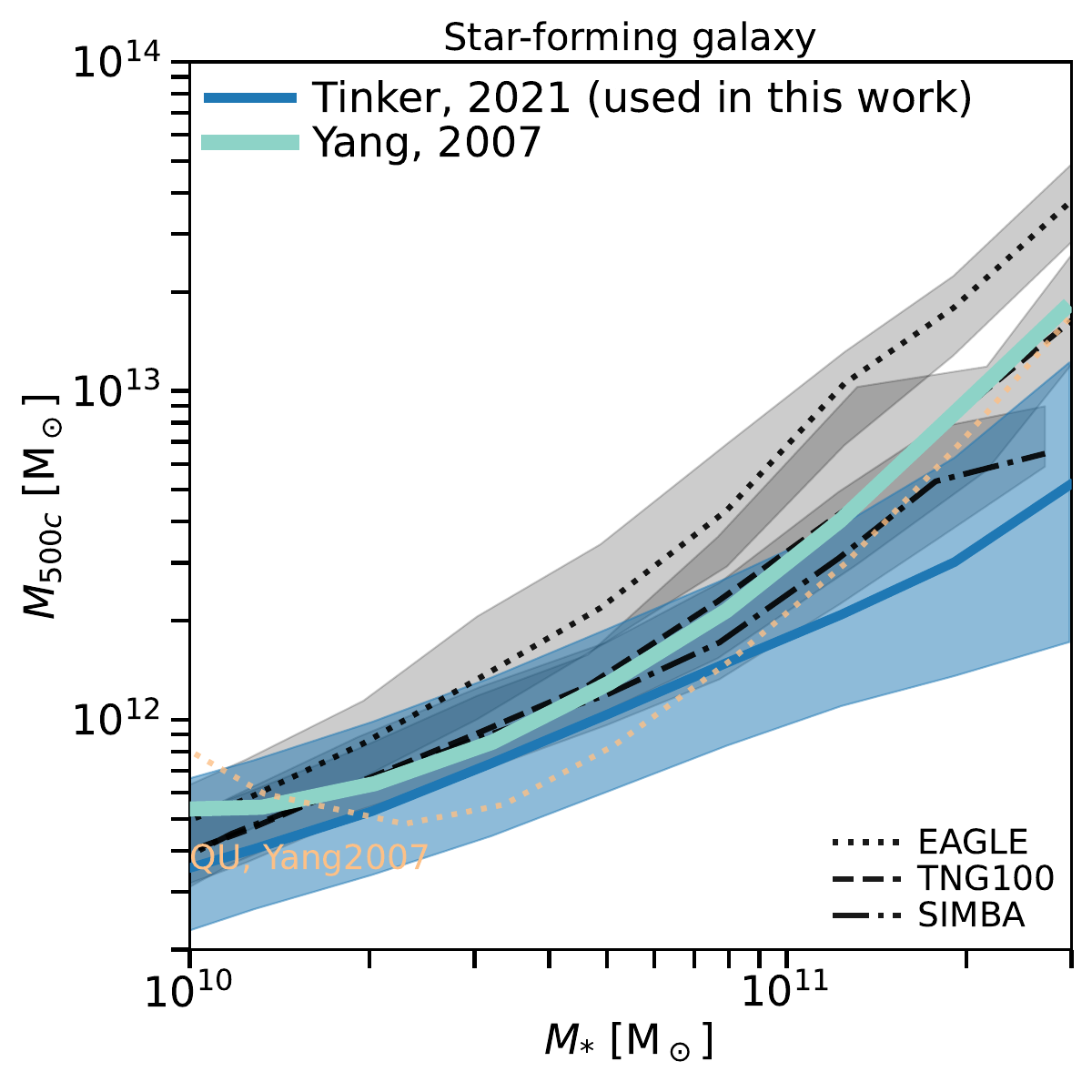}
        \includegraphics[width=0.9\columnwidth]{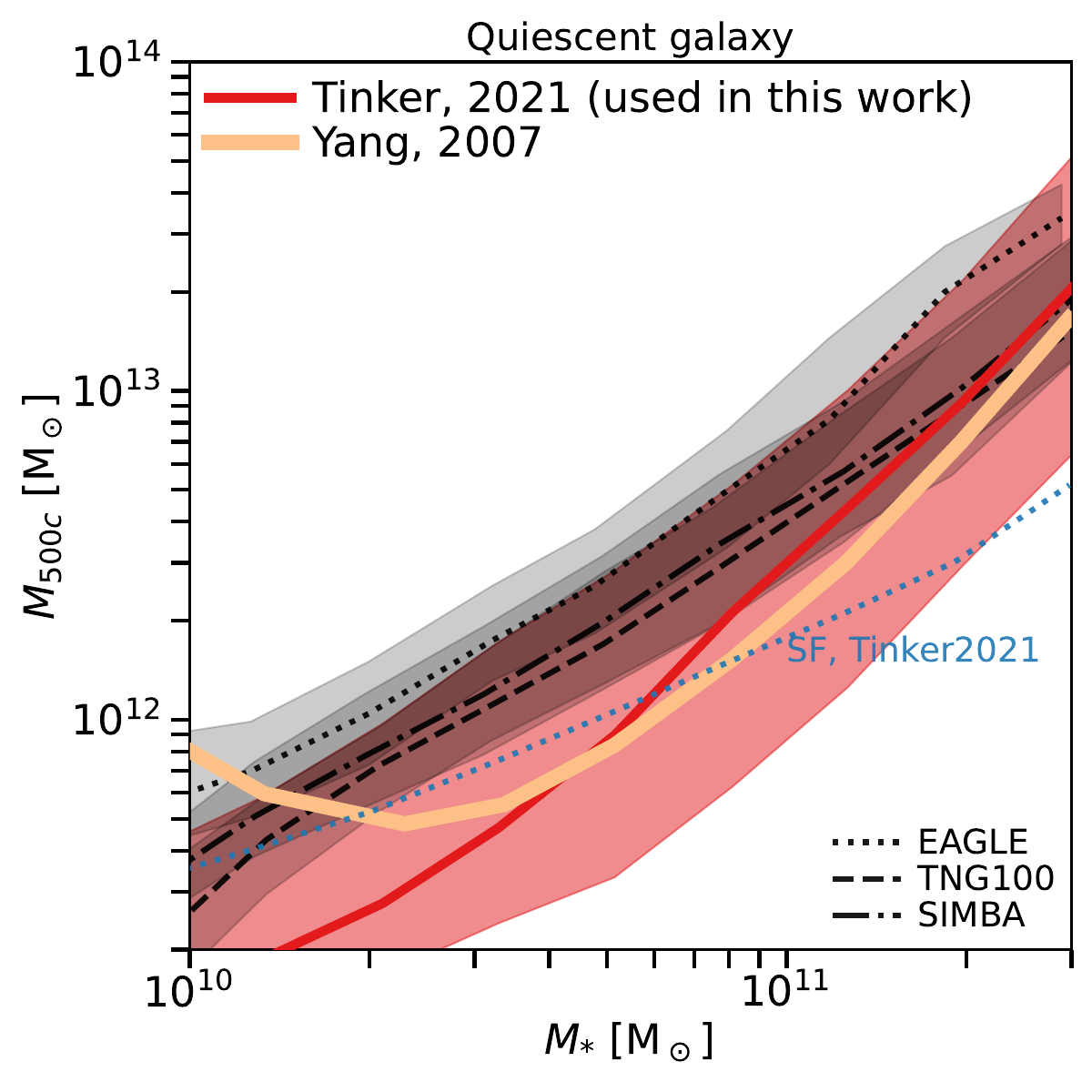}
\caption{Comparison of $M_{\rm 500c}-M*$ relations of central star-forming (top) and quiescent (bottom) galaxies in catalog \citet{Tinker2021} and \citet{Yang2007}, and EAGLE, TNG100 and SIMBA simulations. The shadow area denotes the 16--84\% scatter of the relation. \citet{Yang2007} only includes galaxy groups with $\log(M_{\rm 500c})>11.5$ and causes the turnover of the quiescent SHMR at low mass end.}
        \label{Fig_shmr_yang}
\end{figure}

To figure out if different group finders affect the $L_{\rm X,CGM}$-$M_{\rm 500c}$ scaling relation, we build another CEN$_{\rm halo}$ galaxy sample (we call it CEN$_{\rm halo, Yang}$) using the halo mass provided by \citet{Yang2007}. \citet{Yang2007} assigns halo mass to galaxies using the characteristic luminosity (similar to the total luminosity of the galaxy group), differently, the resulting SHMR is approximately unimodal, that SHMR of star-forming is slightly higher than quiescent galaxies. In Fig.~\ref{Fig_shmr_yang}, we compare the SHMR of star-forming and quiescent galaxies in \citet{Tinker2021} and \citet{Yang2007}. The SHMR of quiescent galaxies is consistent in the two works, while \citet{Yang2007} assigns higher halo mass than \citet{Tinker2021} to star-forming galaxies with $\log(M_*)>10.7$.

\begin{figure}[h!tb]
    \centering
    \includegraphics[width=0.9\columnwidth]{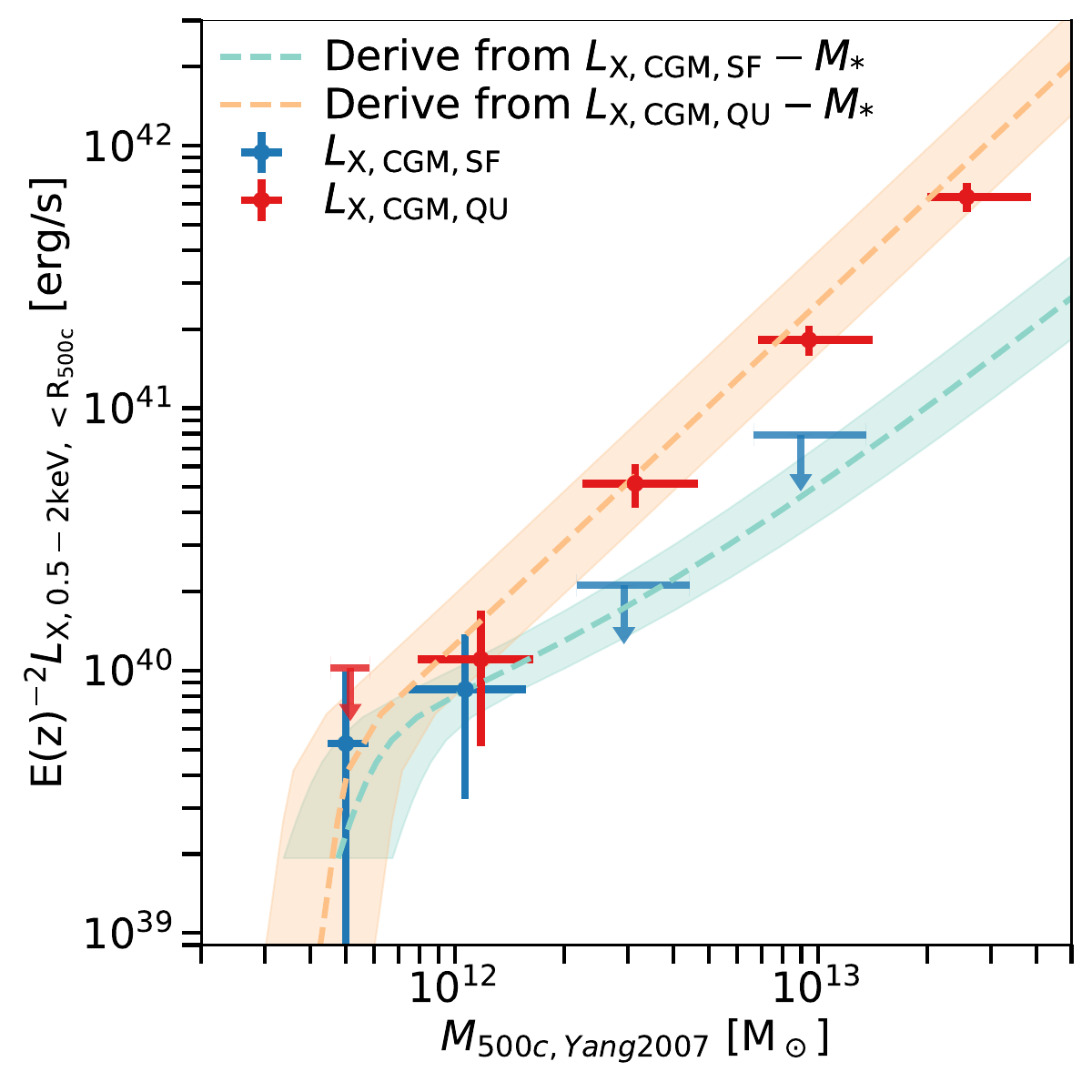}
\caption{Observed hot CGM X-ray luminosity of central star-forming galaxies ($L_{\rm X,CGM, SF}$, blue) and quiescent galaxies ($L_{\rm X,CGM, QU}$, red) selected from \citet{Yang2007}, compared to the derived $L_{\rm X,CGM, SF}$ or $L_{\rm X,CGM, QU}$ by convolving observed $L_{\rm X,CGM}-M_*$ with SHMR of \citet{Yang2007}. The shadow area is the uncertainty from SHMR.}
        \label{Fig_profile_cenhaloyang}
\end{figure}

\begin{figure}[h!tb]
    \centering
    \includegraphics[width=0.9\columnwidth]{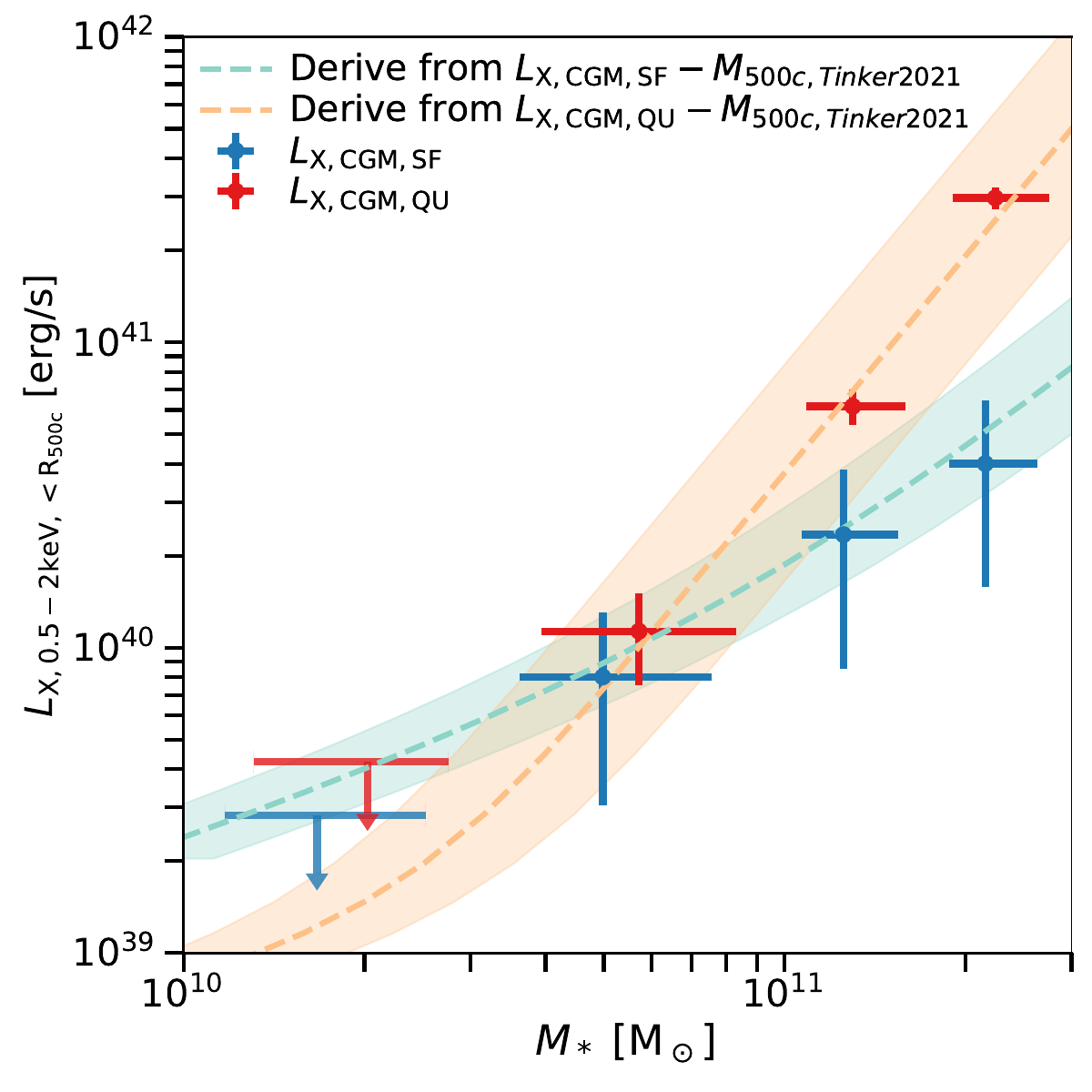}
\caption{Observed $L_{\rm X,CGM, SF}-M_*$ (blue) and $L_{\rm X,CGM, QU}-M_*$ (red) scaling relations, compared to the derived ones by convolving $L_{\rm X, CGM}-M_{\rm 500c}$ with SHMR of \citet{Tinker2021}.}
        \label{Fig_profile_cenhalotinker}
\end{figure}

We follow the procedures in Sect.~\ref{Sec_method} to calculate the $L_{\rm X,CGM}$ of galaxies in the CEN$_{\rm halo, Yang}$ samples. The $L_{\rm X,CGM}$-$M_{\rm 500c, Yang2007}$ scaling relation is plotted in Fig.~\ref{Fig_profile_cenhaloyang}. The star-forming galaxies host fainter X-ray emission than quiescent galaxies, different from the conclusion we drew in Sect.~\ref{Sec_cenhalo} and Fig.~\ref{Fig_profile_cenhalo}. 
The reason is the SHMR used to calibrate the halo mass in \citet{Yang2007}. 
By convolving the observed $L_{\rm X,CGM}$-$M_{*}$ scaling relation with the SHMR used in \citet{Yang2007}, we obtain the derived $L_{\rm X,CGM}$-$M_{\rm 500c, Yang2007}$ (green and orange dashed lines), that is consistent with the observation, as plotted in Fig.~\ref{Fig_profile_cenhaloyang}. Therefore, the $L_{\rm X,CGM}$-$M_{*}$ and $L_{\rm X,CGM}$-$M_{\rm 500c, Yang2007}$ are related by the SHMR, and the approximately unimodal SHMR in \citet{Yang2007} causes the preservation of the similar bifurcation feature in $L_{\rm X,CGM}$-$M_{*}$ and $L_{\rm X,CGM}$-$M_{\rm 500c, Yang2007}$ relations. 
Differently, the bimodal SHMR in \citet{Tinker2021} positions quiescent galaxies in higher-mass dark matter halos, especially for galaxies above $\log(M_*)=10.8$ (see the bottom panel in Fig.~\ref{Fig_shmr_yang}), that compensates the brighter $L_{\rm X, CGM}$ for quiescent galaxies in the high-stellar-mass end, and caused the observed consistent $L_{\rm X,CGM}$-$M_{\rm 500c, Tinker2021}$ scaling relations for star-forming and quiescent galaxies. Indeed, we convolved the observed $L_{\rm X,CGM}$-$M_{\rm 500c, Tinker2021}$ scaling relation with the bimodal SHMR used in \citet{Tinker2021} and obtain consistent $L_{\rm X,CGM}$-$M_{*}$ relations with observation, see Fig.~\ref{Fig_profile_cenhalotinker}.


We conclude that the consistent $L_{\rm X, CGM}$-$M_{\rm 500c}$ scaling relations for star-forming and quiescent galaxies (Fig.~\ref{Fig_profile_cenhalo}) are conditional upon the bimodal SHMR observed by lensing \citep{MandelbaumWangZu_2016MNRAS.457.3200M, Bilicki2021}. Unimodal SHMR relations (such as the one in \citet{Yang2007} ) would give rise to different scaling relations for star-forming and quiescent galaxies. The upcoming galaxy surveys and weak lensing studies would measure more accurate bimodal SHMR, which would help to consolidate our conclusion \citep{DESICollaborationAghamousaAguilar_2016arXiv161100036D,deJong2019,LaureijsAmiauxArduini_2011arXiv1110.3193L,Ivezic_LSST_2019ApJ...873..111I}.

Nevertheless, the dominant importance of halo mass in determining hot CGM X-ray luminosity rhymes with the halo quenching model and the empirical galaxy-halo connection models such as EMERGE and UniverseMachine, where halo mass (or together with the halo assembly history) can solely determine the central galaxy quenching, without introducing the dependence on stellar mass and environment \citep{ZuMandelbaum_2016MNRAS.457.4360Z, BehrooziWechslerHearin_2019MNRAS.488.3143B,Moster2020}. 
Interestingly, the $L_{\rm X,CGM}$-$M_{*}$ relations for star-forming and quiescent galaxies, which are independent of SHMR, bifurcate at about $\log(M_*)=11.0$ ($\log(M_{\rm 200m})\approx12.5$). The bifurcation mass is close to the mass scale of the peak stellar-to-halo-mass ratio and the critical mass above which the virial shocks are stable or the CGM is entirely virialized, $\log(M_*)=10.5$ ($\log(M_{\rm 200m})\approx12.0$) \citep{Dekel2006, Faucher2023}. The similar mass scales suggest a shared underlying physical mechanism, such as the suppression of star formation due to hot accretion. We leave the further modeling and interpretation to future works.



\subsection{Comparison to hydrodynamic simulations}\label{Sec_sim}
\begin{figure*}[h!tb]
    \centering
    \includegraphics[width=0.45\linewidth]{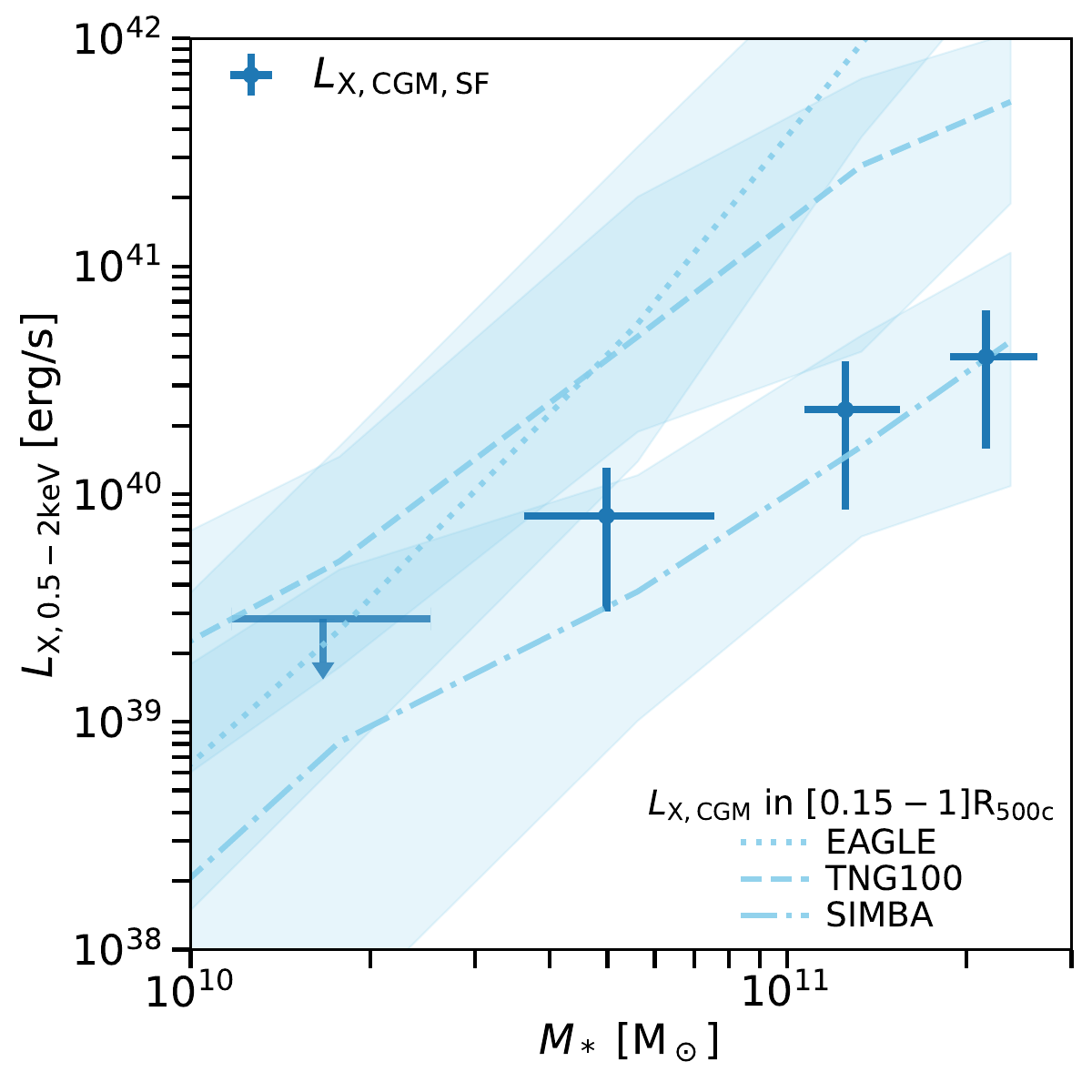}
    \includegraphics[width=0.45\linewidth]
    {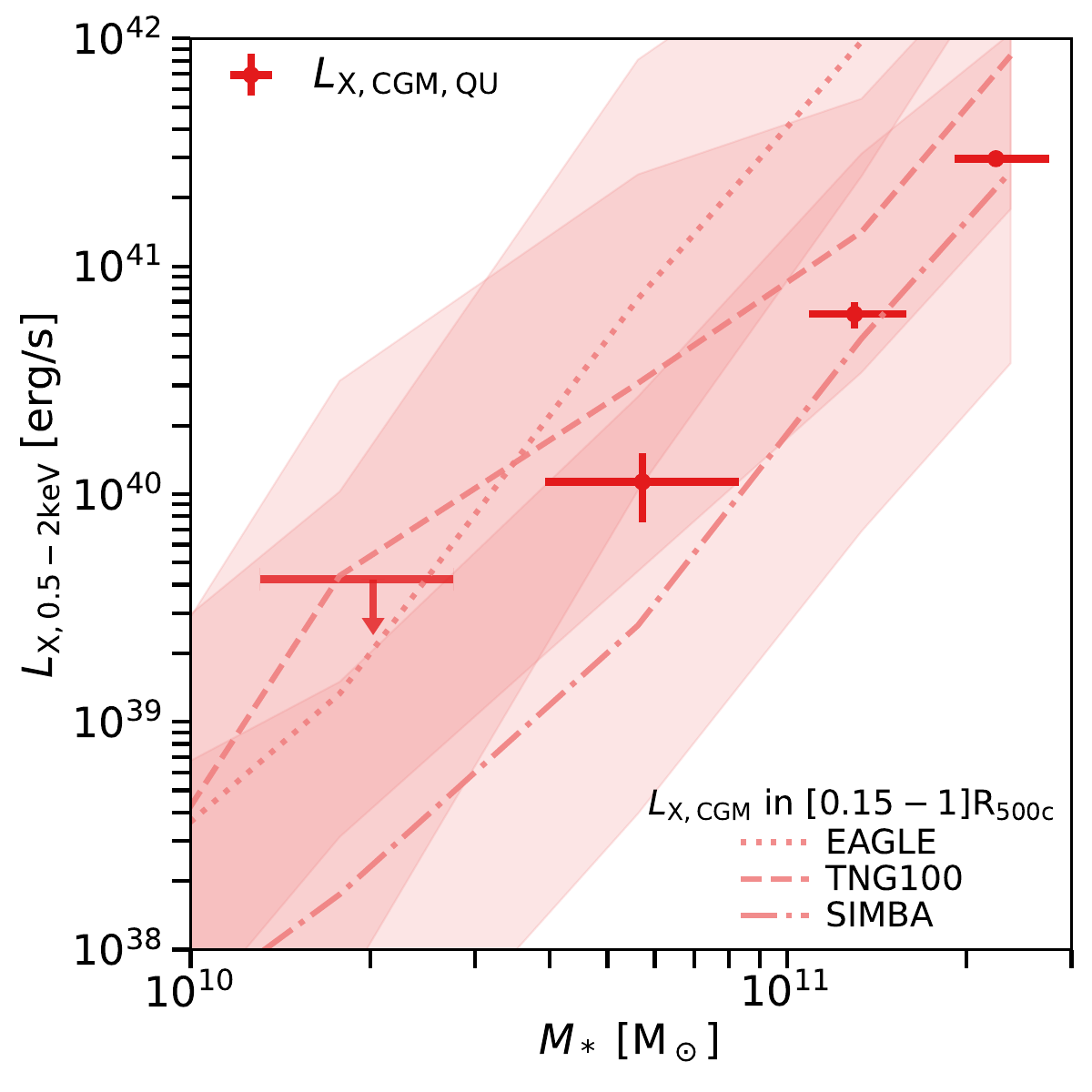}
    \includegraphics[width=0.45\linewidth]
    {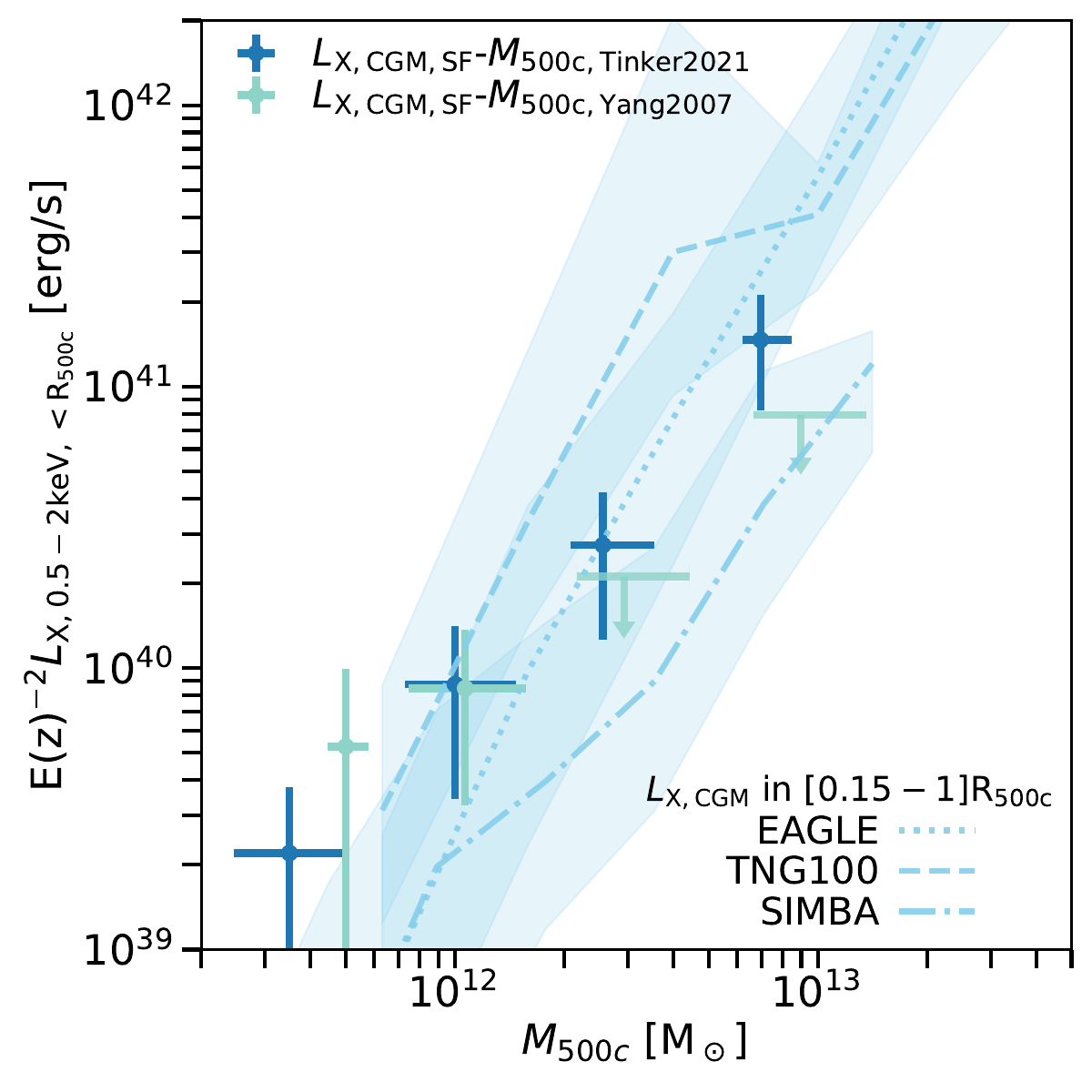}
    \includegraphics[width=0.45\linewidth]{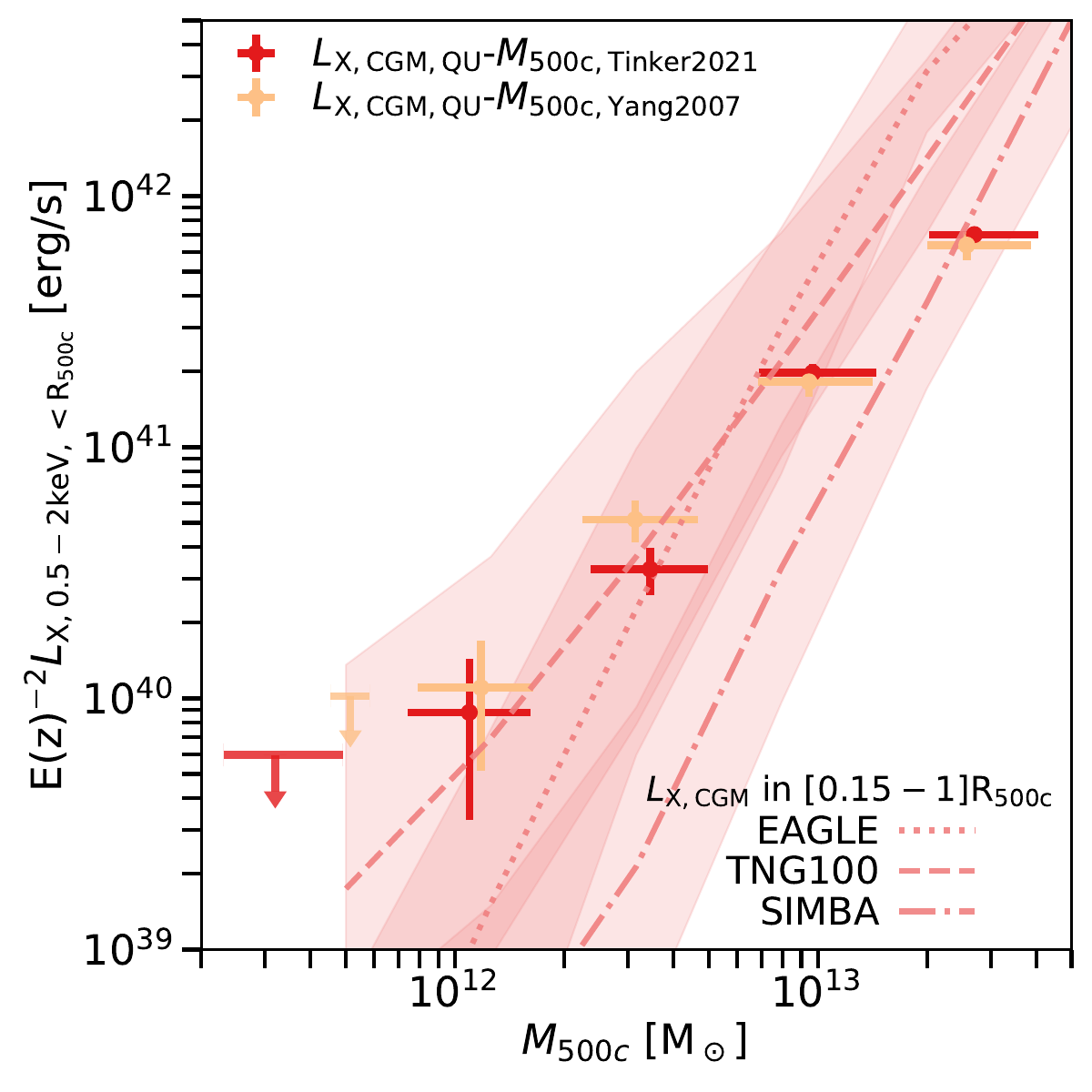}
    \caption{Comparison of the observed $L_{\rm X,CGM}-M_*$ relation to the predictions from the EAGLE, TNG100, and SIMBA simulations with $1\sigma$ uncertainties, for star-forming (top left) and quiescent (top right) galaxies. Comparison of the observed $L_{\rm X,CGM}-M_{\rm 500c}$ relation to the predictions from the EAGLE, TNG100, and SIMBA simulations with $1\sigma$ uncertainties, for star-forming (bottom left) and quiescent (bottom right) galaxies.}
        \label{Fig_sim}
\end{figure*}

We compare the observed $L_{\rm X,CGM}-M_{\rm *}$ and $L_{\rm X,CGM}-M_{\rm 500c}$ scaling relations to the hydrodynamical simulations.
We use three cosmological hydrodynamical simulations: EAGLE \citep{Crain2015,Schaller2015,Schaye2015,McAlpine2016}, TNG100 from IllustrisTNG \citep{Marinacci2018,Naiman2018,Springel2018,Pillepich2018,Nelson2019a,Nelson2019b}, and SIMBA \citep{Dave2019}. 
For each simulated galaxy, the X-ray luminosity of the CGM is measured in the $0.5-2$ keV range and within a radial range of $(0.15-1)\,R_{\rm 500c}$ \citep{Truong2021b, Comparat2022}, the stellar mass is calculated within twice the half-stellar mass radius, and $M_{500c}$ is the total mass within $R_{\rm 500c}$. 
The 4000\AA \ break definition of star-forming and quiescent galaxies is hard to reproduce in the simulations. However, as shown in Sect.~\ref{Sec_sfr}, the 4000\AA \ break definition gives consistent results with the specific SFR (sSFR) $=10^{-11}\rm yr^{-1}$ separation.
We take the instantaneous SFR within twice the stellar half-mass radius of simulated galaxies to calculate sSFR and separate star-forming and quiescent galaxies.

We select and bin simulated star-forming and quiescent galaxies in the same $M_*$ bins of the CEN$_{\rm SF}$ and CEN$_{\rm QU}$ samples, and the same $M_{\rm 200m}$ bins of the CEN$_{\rm halo, SF}$ and CEN$_{\rm halo, QU}$ samples. We calculate the mean $L_{\rm X,CGM}$ and take its uncertainty as the $16-84\%$ scatter of the luminosity of the galaxies. The uncertainty reflects the approximately 1-dex galaxy-to-galaxy variation across most of the mass range of simulated galaxies; notice it is different from the uncertainty defined in the observation that only reflects the variation of the mean X-ray luminosity of certain galaxy populations.
We emphasize that the comparisons between observations and simulations in this work are made at face value, as we do not create a dedicated mock catalog to reproduce the stellar mass function, halo mass function and SHMR of CEN$_{\rm SF}$, CEN$_{\rm QU}$, CEN$_{\rm halo, SF}$ and CEN$_{\rm halo, QU}$ samples. 

We compare the simulation predicted $L_{\rm X, CGM}-M_*$ to observation in the top row of Fig.\ref{Fig_sim}. The different feedback models in the three simulations result in different predictions on the hot CGM properties \citep{Truong2023, Wright2024}. 
EAGLE predicts consistent X-ray emission at the low-mass end but overpredicts the CGM emissions for star-forming and quiescent galaxies with $\log(M_*)>11.0$ by about 1-dex. This discrepancy suggests that the feedback model in EAGLE may be less efficient at removing gas around massive galaxies \citep{Davies2019}.
TNG100 predicts brighter X-ray emission around star-forming galaxies with $\log(M_*)>10.5$ and consistent X-ray emission around quiescent galaxies, which might suggest the feedback induced gas depletion for star-forming galaxies is less efficient in TNG100 \citep{Truong2020}.
SIMBA predicts consistent X-ray emission around star-forming and quiescent galaxies at the high-mass end but underestimates X-ray emission at the low-mass end. More dedicated comparisons with simulations are essential to interpret the underlying physics, such as the interplay between stellar and AGN feedback \citep{Medlock2024}.  

We compare the $L_{\rm X,CGM}-M_{\rm 500c}$ relations predicted by the simulations to the observation based on halo mass provided by \citet{Tinker2021} and \citet{Yang2007} in the bottom row of Fig.~\ref{Fig_sim}.
Differently from the $L_{\rm X, CGM}-M_*$ case, we notice EAGLE predicts consistent X-ray emission around star-forming galaxies with $M_{\rm 500c, Tinker2021}$, but still overpredicts $L_{\rm X, CGM}$ with $M_{\rm 500c, Yang2007}$. 
The reason is the same as discussed in Sect.~\ref{Sec_lxm}, the different SHMR realized in EAGLE, as compared in Fig.\ref{Fig_shmr_yang}. The EAGLE simulation results in almost 1-dex higher halo mass for star-forming galaxies than \citet{Tinker2021}, which compensates for the overpredicted $L_{\rm X, CGM}-M_*$ relation. 
The difference between the star-forming SHMR of EAGLE and \citet{Yang2007} is smaller but still exists, therefore the overprediction is maintained. 
Star-forming SHMR of TNG100 agrees with \citet{Yang2007} but is steeper than \citet{Tinker2021}, as a result, TNG100 prediction is higher than $L_{\rm X,CGM}-M_{\rm 500c, Yang2007}$ but consistent with $L_{\rm X,CGM}-M_{\rm 500c,Tinker2021}$.
SIMBA predicts consistent star-forming SHMR with \citet{Yang2007} and slightly steeper SHMR than \citet{Tinker2021}, and correspondingly consistent $L_{\rm X,CGM}-M_{\rm 500c,Yang2007}$ and lower $L_{\rm X,CGM}$ than the one measured in \citet{Tinker2021}'s halo mass.
Regarding quiescent galaxies, the three simulations predict consistent SHMR above $\log(M_*)=11.0$ and higher halo mass for low-mass quiescent galaxies. Correspondingly, EAGLE and TNG100 predict consistent X-ray emission, and SIMBA significantly underestimates the X-ray emission around quiescent galaxies.

Notice though SIMBA predicts that quiescent galaxies host brighter hot CGM, TNG100 and EAGLE do not show a significant difference between star-forming and quiescent; it’s important to emphasize that this result applies only when considering the sample mean average. When considering the sample median average, all three simulations predict that star-forming galaxies are comparable or more X-ray luminous than their quiescent counterparts, as shown in previous studies \citep{Truong2020,OppenheimerBogdanCrain_2020ApJ...893L..24O,Truong2023}. This discrepancy can be explained by the fact that the mean X-ray emission is likely biased by a small number of highly luminous galaxies in the simulation. Indeed, it is intrinsically difficult to reproduce the observed scatter of SHMR or $L_{\rm X, CGM}-M_*$ relation in simulations, limited by the simulation volumes and number of galaxies. 
In summary, the different SHMR used in group finders and simulations complicates the interpretation of our comparison. The discrepancies in the $L_{\rm X,CGM}-M_{\rm 500c}$ or $L_{\rm X, CGM}-M_*$ relationships could arise from both the underlying feedback physics, the methodologies used to estimate $M_{\rm 500c}$ and the intrinsic scatter of galaxy population. 
A joint and model-forward analysis of SHMR, $L_{\rm X,CGM}-M_{\rm *}$ and $L_{\rm X,CGM}-M_{\rm 500c}$ for star-forming and quiescent galaxies in the three simulations is necessary, which we leave for future work.

We emphasize that the SHMR in the simulation is usually calibrated to agree with observation. Indeed at lower(higher) mass, the star-forming (quiescent) SHMR of the three simulations agrees within dispersion with \citet{Yang2007} and \citet{Tinker2021} (see Fig.~\ref{Fig_shmr_yang}). Considering star-forming (quiescent) galaxies are dominant at low-mass (high-mass), the average SHMR of the simulations before splitting star-forming and quiescent agrees with observation \citep{Wright2024}.

\subsection{Definition of star-forming and quiescent galaxies}\label{Sec_sfr}

\begin{figure}[h!tb]
    \centering
    \includegraphics[width=0.9\columnwidth]{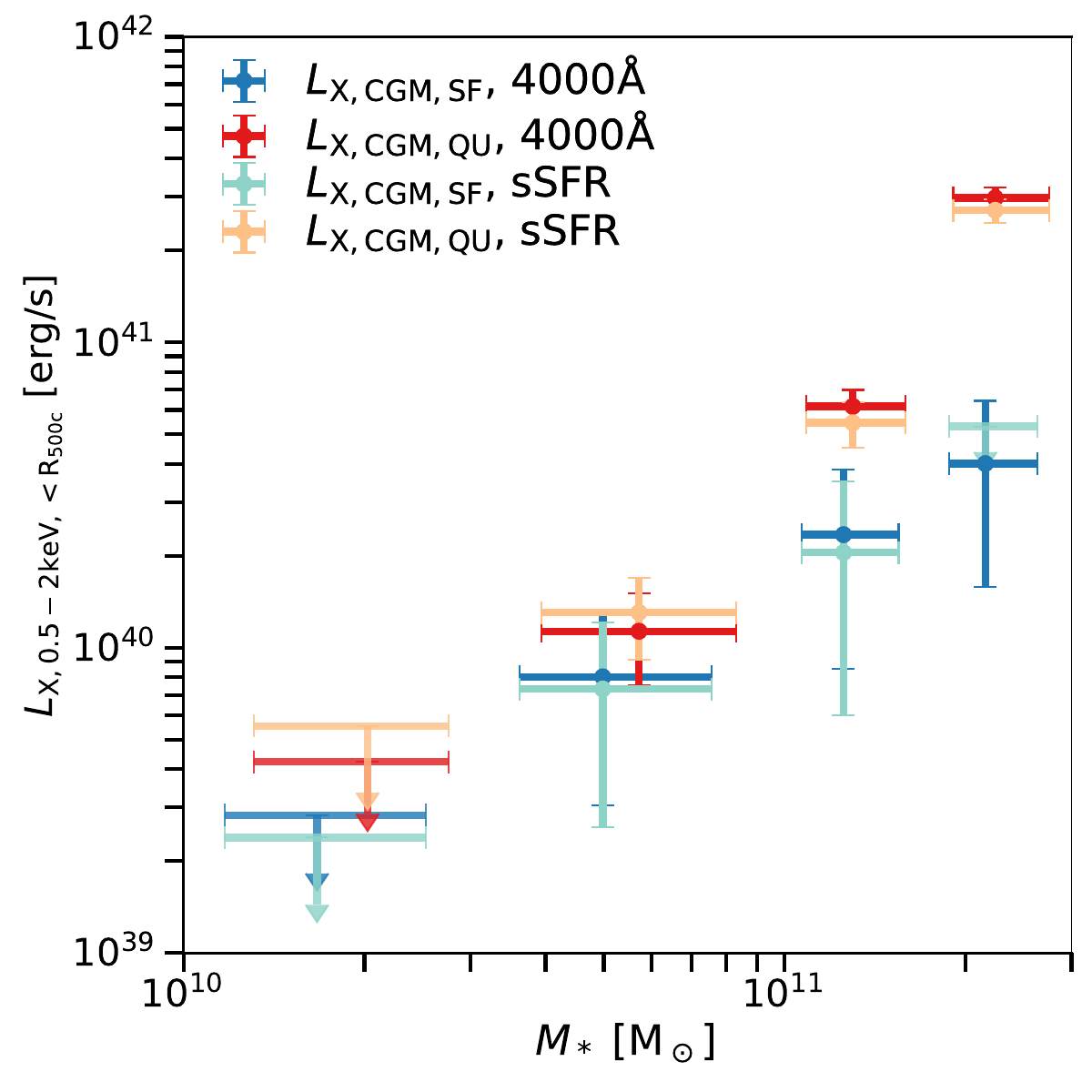}
\caption{Comparison the $L_{\rm X,CGM}-M_*$ relations of star-forming and quiescent galaxies selected by 4000\AA \ break or sSFR.}
        \label{Fig_lxm_ssfr}
\end{figure}

We used the 4000\AA \ break to separate star-forming and quiescent galaxies and measured the $S_{\rm X, CGM}$ and $L_{\rm X, CGM}$ in Sect.~\ref{Sec_result}. The 4000\AA \ break is hard to reproduce in simulations. Besides, we want to verify our results are not sensitive to the specific selection of star-forming and quiescent galaxies.

As a test, we define the star-forming and quiescent galaxies by the broadly-used sSFR: galaxies with sSFR$<10^{-11}\rm yr^{-1}$ are quiescent, and galaxies with sSFR$>10^{-11}\rm yr^{-1}$ are star-forming. We follow the procedure in Sect.~\ref{Sec_method} to calculate the $L_{\rm X,CGM,sSFR}$ for the new star-forming and quiescent samples, the resulted $L_{\rm X,CGM, sSFR}-M_*$ scaling relation is compared to the 
one of Sect.~\ref{Sec_cen} in Fig.~\ref{Fig_lxm_ssfr}. We find consistent results and verify the definition of star-forming and quiescent galaxies does not change our results.

\subsection{Comparison with previous studies}\label{Sec_lit}

Using the PV/eFEDS data covering a sky area of $140\ \rm deg^2$, two papers studied the hot CGM around star-forming and quiescent galaxies by stacking eROSITA X-ray data \citep{Comparat2022, Chada2022}. In this section, we compare our results to them and explain the factors that caused the difference.

\begin{figure*}[h!tb]
    \centering
    \includegraphics[width=0.45\linewidth]
    {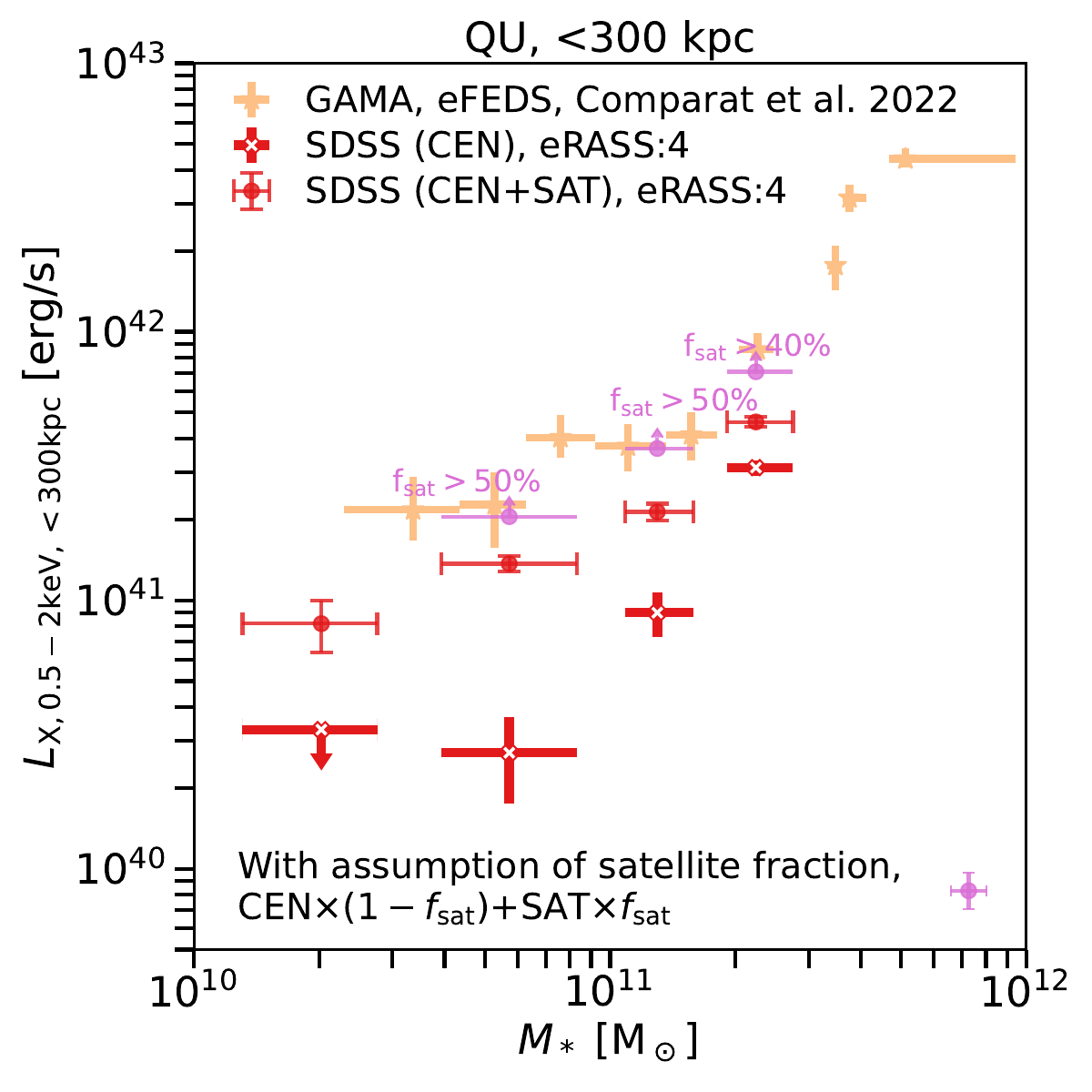}
    \includegraphics[width=0.45\linewidth]{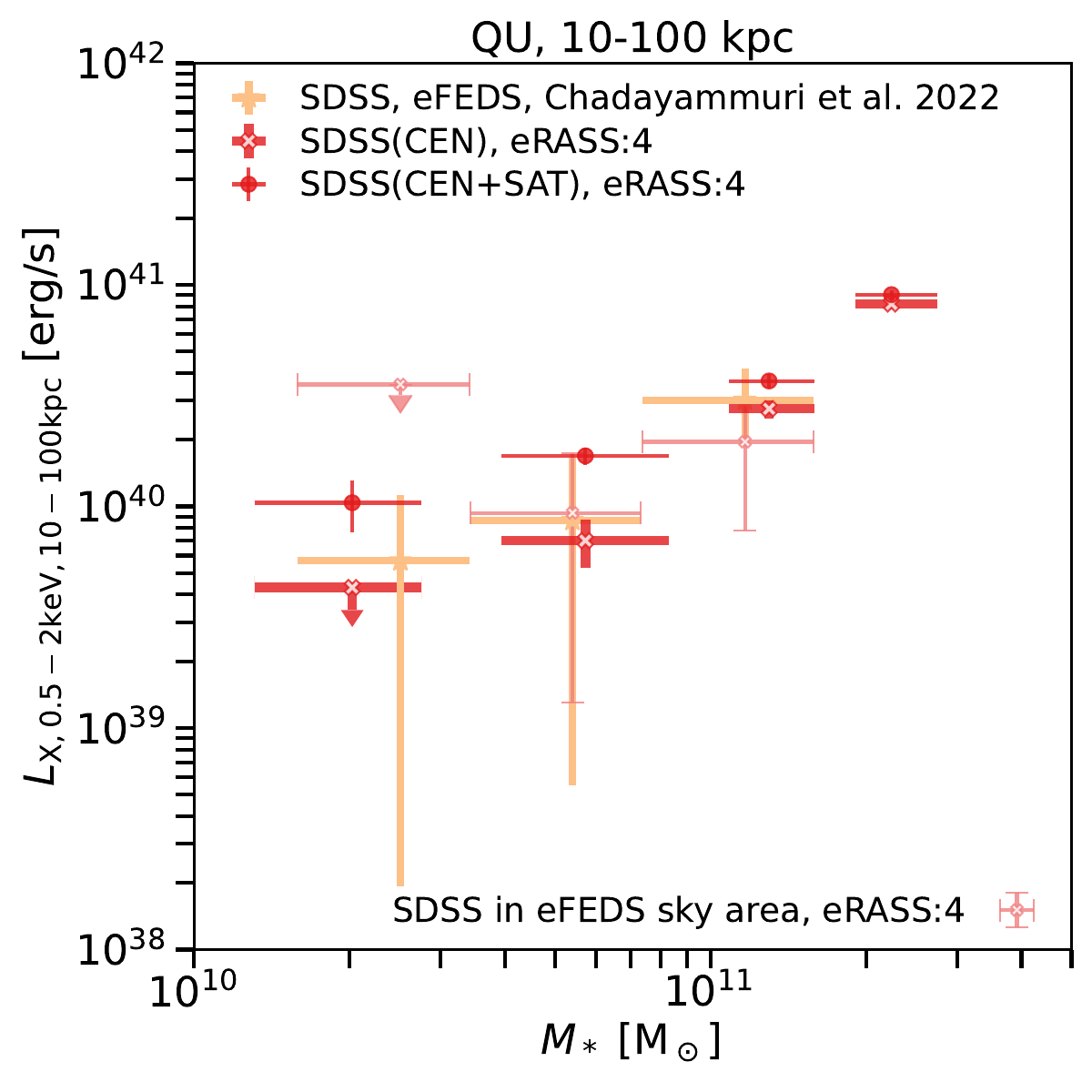}
    \includegraphics[width=0.45\linewidth]{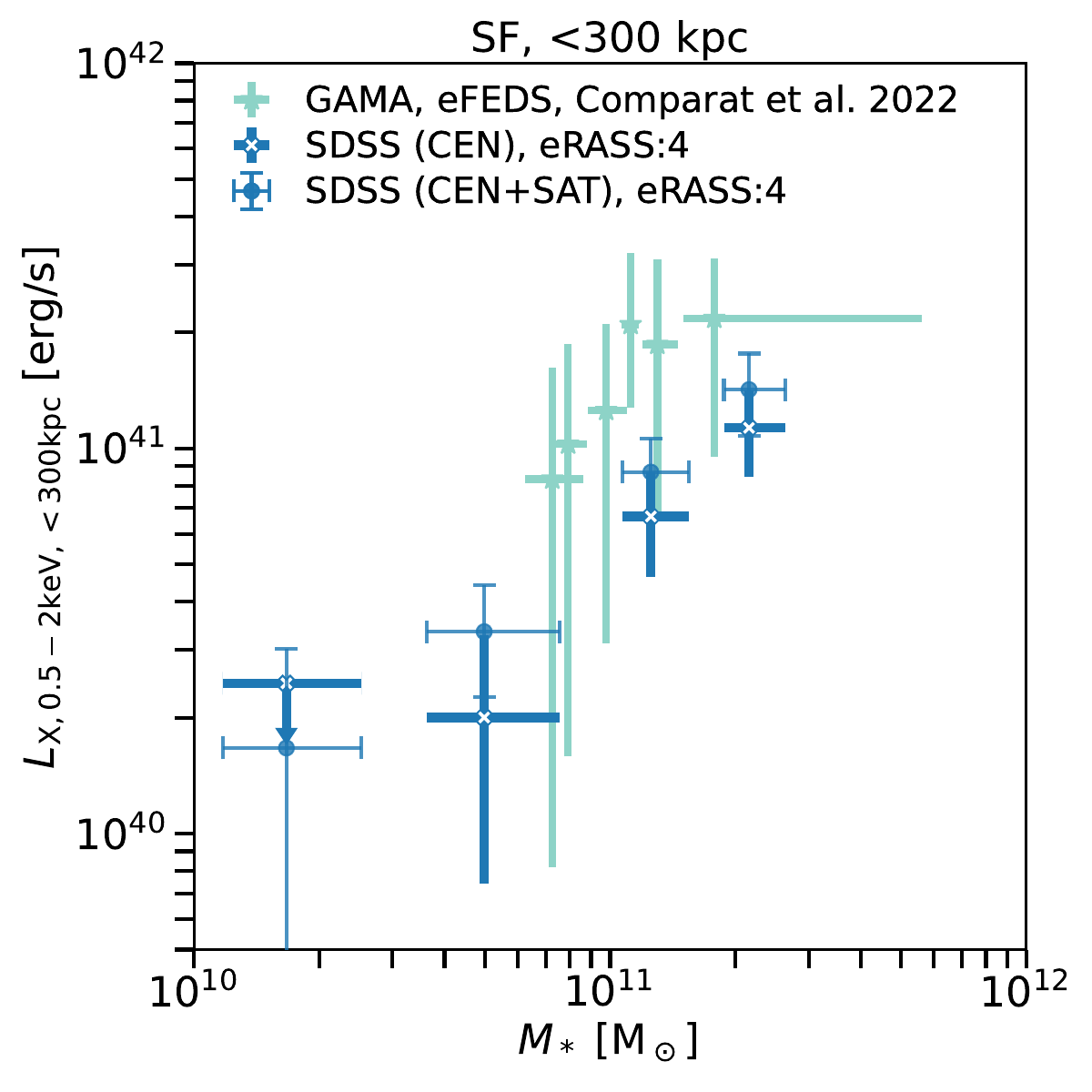}
    \includegraphics[width=0.45\linewidth]{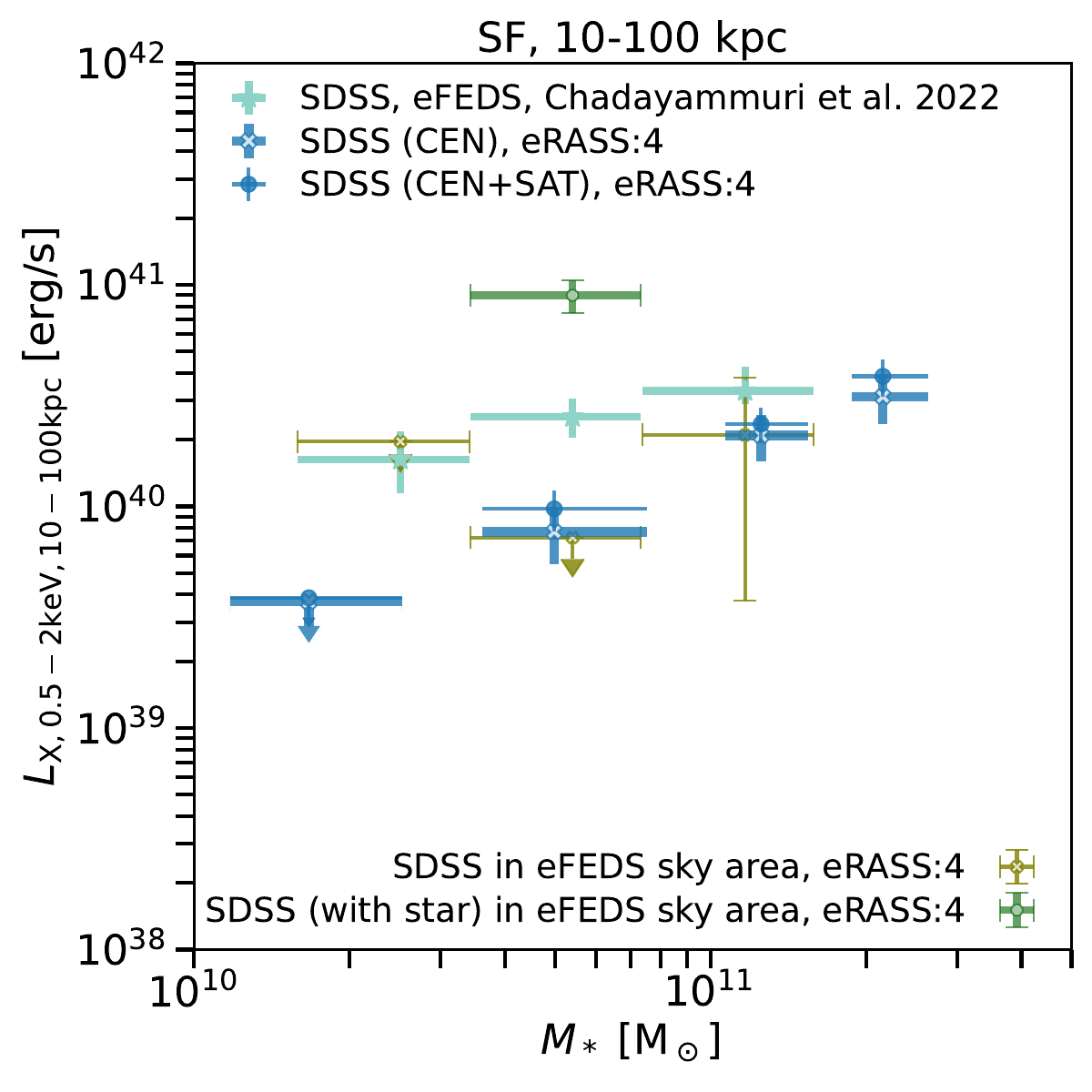}
    \caption{X-ray luminosity within 300 kpc, without masking extended X-ray sources, for quiescent (top left) and star-forming (bottom left) central galaxies, central and satellite galaxies together, compared to \citet{Comparat2022}. Also shown are the satellite fractions to explain the bright $L_{\rm X}$ of quiescent galaxies \citet{Comparat2022}. X-ray luminosity within $10-100$ kpc, after masking all X-ray sources and subtracting XRB emission, for quiescent (top right) and star-forming (bottom right) central galaxies, central and satellite galaxies together, compared to \citet{Chada2022}. Also shown is the $L_{\rm X, CGM}$ of central star-forming galaxies selected in eFEDS sky area and galaxies with star companion.}
        \label{Fig_compare}
\end{figure*}

\subsubsection{Comparison to Comparat et al. (2022)}

\citet{Comparat2022} stack 16142 `central' galaxies at $0.05\le z_{\rm spec} \le0.3$ taken from the GAMA spectroscopic survey. They define the `central' galaxies as the most massive galaxy within $2R_{\rm vir}$ and found about 90\% galaxies in GAMA are central. 
During stacking, the X-ray-detected galaxy clusters associated with the stacked galaxies are not masked. 
They find that quiescent galaxies' X-ray surface brightness profiles are brighter and more extended than star-forming galaxies. They provide the scaling relation between X-ray luminosity within 300 kpc ($L_{\rm X,<300kpc}$ and the stellar mass of galaxies (Fig. 7 in \citet{Comparat2022}).

We compare our stacking based on our CEN$_{\rm SF}$ and CEN$_{\rm QU}$ samples to \citet{Comparat2022} in the left column of Fig.~\ref{Fig_compare}. To keep consistency, we do not mask extended X-ray sources and integrate the X-ray emission within 300 kpc around galaxies. We obtain much lower $L_{\rm X,<300kpc}$ of quiescent galaxies than \citet{Comparat2022}. Except for the reasons of different source masking and galaxy samples, the most plausible reason for the discrepancy is the satellite boost bias in \citet{Comparat2022}. In \citet{Zhang2024profile}, we stack central and satellite galaxies respectively and find more extended and brighter X-ray emission around satellite galaxies, which is due to the contamination from the nearby massive galaxies around stacked satellites. Consequently, including the satellite galaxies (misclassified as central) in the galaxy sample would boost the X-ray emission (See also \citet{Shreeram2024}, where they quantify this effect and conclude that for $\log(M_*)\approx 11.0$ galaxies, the satellite boost effect cannot be ignored). 
To remove this kind of satellite boost bias, we stack only central galaxies in our study, where the central galaxies are identified by the group finder with high accuracy. The central galaxy fraction of CEN$_{\rm QU}$ is about $70$\%.
If we include also satellite galaxies in the stacking, the $L_{\rm X,<300kpc}$ would increase by about 2--3 times, as shown in the upper left panel of Fig.~\ref{Fig_compare}, which is, however, still lower $L_{\rm X,<300kpc}$ than \citet{Comparat2022}.
Considering the GAMA survey ($r_{\rm AB}=19.8$) is deeper than SDSS ($r_{\rm AB}=17.8$), the central galaxy fraction of GAMA is expected to be lower, as more fainter galaxies, namely potential satellite galaxies, are included in the sample. 
We assume the mean X-ray luminosity of satellite (and central) galaxies in the GAMA catalog has the same bright X-ray emission as SDSS, and the bright $L_{\rm X,<300kpc}$ of \citet{Comparat2022} is solely due to the satellite boost bias. 
We scale the satellite fraction ($f_{\rm sat}$) and the corresponding $L_{\rm X,<300kpc}$ to the $L_{\rm X,<300kpc}$ values from \citet{Comparat2022}. We obtain the satellite fraction lower-limit of the GAMA galaxy sample used in \citet{Comparat2022}. We find a satellite fraction of about 40--50\% can explain the bright $L_{\rm X,<300kpc}$ in \citet{Comparat2022}.

We find consistent $L_{\rm X,<300kpc}$ of star-forming galaxies with \citet{Comparat2022}. This is because, first, the X-ray emission around star-forming galaxies is less extended and bright, therefore less contaminating nearby galaxies. Second, the central fraction of star-forming galaxies is high (about 80\%). As a result, star-forming galaxies are less affected by the satellite boost bias.

\subsubsection{Comparison to Chadayammuri et al. (2022)}
\citet{Chada2022} stacked about 1600 galaxies at $0.01<z_{\rm spec}<0.1$ selected from 2643 galaxies located in the overlap sky area of SDSS and eFEDs\citep{Ahn2014,Montero2016}. They do not select central galaxies but mask all known X-ray-detected galaxy clusters, in this way, most satellite galaxies (and part of central galaxies) are excluded. They obtain slightly brighter or comparable X-ray emission around star-forming galaxies than quiescent galaxies. They provide the scaling relation between X-ray luminosity within 10--100 kpc ($L_{\rm X,10-100 kpc}$) and the stellar mass of galaxies (Fig. 4 in \citet{Chada2022}).

We calculate the X-ray luminosity within $10-100$ kpc of CEN$_{\rm SF}$ and CEN$_{\rm QU}$ samples and compare them to \citet{Chada2022} in the right column of Fig.~\ref{Fig_compare}. We find consistent $L_{\rm X,10-100 kpc}$ of quiescent galaxies with \citet{Chada2022}, and lower $L_{\rm X,10-100 kpc}$ in the star-forming galaxies than \citet{Chada2022}. 
To figure out the reasons for the difference in the star-forming galaxies, we stack the galaxies in CEN$_{\rm SF}$ sample located in the eFEDS region. We find consistent $L_{\rm X,10-100kpc}$ with \citet{Chada2022} for the lowest stellar mass bin. This suggests that a limited number of galaxies (about 400) located in the small volume may not be representative of the average properties of the galaxy population. Indeed, an overabundance of dwarf star-forming galaxies is found in the eFEDS area \citep{Vulic2022}. 
The bright $L_{\rm X,10-100kpc}$ of the middle stellar mass bin in \citet{Chada2022} is also inconsistent with our results. We observe this might be due to the star contamination in the SDSS galaxy sample in \citet{Chada2022}. Indeed, we find about 175 galaxies in \citet{Chada2022} are not included in \citet{Tinker2021}, within which about 26 (19) star-forming (quiescent) galaxies near bright stars \citep[\texttt{ILSS}=-1,][]{Blanton2005}. We stack the galaxies not included in \citet{Tinker2021} at the middle star-forming stellar mass bin (29 galaxies) and find $L_{\rm X,10-100kpc}\approx10^{41}\rm erg/s$, which can explain the higher $L_{\rm X,10-100kpc}$. For the other star-forming and quiescent stellar mass bins, the star contamination is not prominent after averaging.

We conclude that the accuracy of the stacking result relies on the cleanness and completeness of the central galaxy sample.
By minimizing the satellite boost bias, defining a large enough approximately volume-limited galaxy sample, and carefully masking and modeling point sources, our work provides a more accurate view of the $L_{\rm X}-M_*$ scaling relations for star-forming and quiescent central galaxies at low redshift. 




\section{Conclusions}
\label{Sec_summary}

In this work, we used the X-ray data from the eROSITA all-sky survey to detect the difference in the hot circumgalactic medium around star-forming and quiescent galaxies. We built approximately volume-limited central star-forming and quiescent galaxy samples with stellar mass $\log(M_*)=10.0-11.5$ or halo mass $\log(M_{\rm 200m})=11.5-14.0$ from SDSS DR7 catalog (Table~\ref{Tab:gal:samples} and Fig.~\ref{Fig_SHMR}). We stacked the X-ray emission around galaxies and we report the following findings:
\begin{itemize}
\item We detect extended X-ray emission from hot CGM out to $R_{\rm 500c}$ around star-forming with $\log(M_*)>11.0$ and quiescent galaxies with $\log(M_*)>10.5$. The hot CGM X-ray surface brightness ($S_{\rm X, CGM}$) profile of quiescent galaxies can be described by a $\beta$ model with $\beta \approx 0.4$ (Fig.~\ref{Fig_profile_cen} and Table~\ref{tab:betamodel}). 
\item We measure the scaling relation of the integrated hot CGM luminosity ($L_{\rm X,CGM}$) within $R_{\rm 500c}$ with $M_*$. Above $\log(M_*)=11.0$, $L_{\rm X,CGM}$ of quiescent galaxies is brighter than star-forming galaxies, comparable $L_{\rm X,CGM}$ is found below $\log(M_*)=11.0$ (Fig.~\ref{Fig_profile_cen} and Table~\ref{table:LX:M}).
\item We detect extended X-ray emission in haloes with $\log(M_{\rm 200m})>12.5$ around star-forming and quiescent galaxies out to $R_{\rm 500c}$. The $S_{\rm X, CGM}$ profile can be described by a $\beta$ model with $\beta \approx 0.4-0.5$ (Fig.~\ref{Fig_profile_cenhalo} and Table~\ref{tab:betamodel}).
\item Based on the bimodal stellar-to-halo mass relation (SHMR, \citet{Tinker2021}), we find consistent $L_{\rm X,CGM}-M_{\rm 500c}$ scaling relations for star-forming and quiescent galaxies, which suggests halo mass is the determining factor for the CGM heating (Fig.~\ref{Fig_profile_cenhalo} and Table~\ref{table:LX:M}). 
\item The observed $L_{\rm X,CGM}-M_{\rm 500c}$ relation depends on the SHMR used to estimate the halo mass of galaxies, namely, star-forming galaxies may host less bright CGM than quiescent galaxies if a unimodal SHMR \citep{Yang2007} is adopted (Fig.~\ref{Fig_shmr_yang} and Fig.~\ref{Fig_profile_cenhaloyang}). However, either selected in stellar mass or halo mass and independent of the SHMR, the star-forming galaxies do not host brighter X-ray emission from the hot CGM than their quiescent counterparts.
\item We compare our results to the cosmological hydrodynamical simulations EAGLE, TNG100, and SIMBA (Fig.~\ref{Fig_sim}). We find discrepancies in some stellar or halo mass bins, which might arise from the methodologies used to estimate halo mass, intrinsic scatter of galaxy population and the underlying feedback physics. 
\item We discuss the results of \citet{Comparat2022} and \citet{Chada2022} (Fig.~\ref{Fig_compare}). We find satellite boost bias, a limited number of galaxies, and star contamination contribute to the difference between \citet{Comparat2022} and \citet{Chada2022}. By building large and well-defined central galaxy samples, we minimize all the biases and provide more accurate measurements of the CGM around star-forming and quiescent galaxies.

\end{itemize}
\noindent
We discuss how the $L_{\rm X,CGM}-M_{\rm 500c}$ scaling relations rely on the SHMR used to calibrate the halo mass in group finders and cosmological simulations. The upcoming galaxy surveys and weak lensing studies will provide more accurate SHMR measurements, for example, DESI \citep{DESICollaborationAghamousaAguilar_2016arXiv161100036D}, 4MOST\citep{deJong2019}, Euclid \citep{LaureijsAmiauxArduini_2011arXiv1110.3193L}, LSST \citep{Ivezic_LSST_2019ApJ...873..111I}. Examining whether the hot CGM temperature (or even metallicity) differs or aligns between star-forming and quiescent galaxies will constrain the heating-cooling processes and baryon budget, which provide additional constraints on the feedback and quenching mechanism. We will investigate this through eROSITA X-ray spectra analysis, and future X-ray microcalorimeters \citep{XRISM2020, ZhangLi2022, Barret2023, Schellenberger2024}.
Our measurements provide new bricks to study the quenching mechanism and the role of different feedback processes, either through the semi-analytic models \citep{Singh2024,Voit2024}, or cosmological simulations \citep{Braspenning2024, Vladutescu-Zopp2024, Sultan2024}.

\begin{acknowledgements}

We thank the anonymous referee for thoughtful comments that improved the manuscript. We thank Dominique Eckert for the meaningful discussion. This project acknowledges financial support from the European Research Council (ERC) under the European Union's Horizon 2020 research and innovation program HotMilk (grant agreement No. 865637). GP acknowledges support from Bando per il Finanziamento della Ricerca Fondamentale 2022 dell'Istituto Nazionale di Astrofisica (INAF): GO Large program and from the Framework per l'Attrazione e il Rafforzamento delle Eccellenze (FARE) per la ricerca in Italia (R20L5S39T9). NT acknowledges support from NASA under award number 80GSFC21M0002. PP has received funding from the European Research Council (ERC) under the European Union's Horizon Europe research and innovation program ERC CoG (Grant agreement No. 101045437).\\

This work is based on data from eROSITA, the soft X-ray instrument aboard SRG, a joint Russian-German science mission supported by the Russian Space Agency (Roskosmos), in the interests of the Russian Academy of Sciences represented by its Space Research Institute (IKI), and the Deutsches Zentrum für Luft- und Raumfahrt (DLR). The SRG spacecraft was built by Lavochkin Association (NPOL) and its subcontractors, and is operated by NPOL with support from the Max Planck Institute for Extraterrestrial Physics (MPE).\\

The development and construction of the eROSITA X-ray instrument was led by MPE, with contributions from the Dr. Karl Remeis Observatory Bamberg \& ECAP (FAU Erlangen-Nuernberg), the University of Hamburg Observatory, the Leibniz Institute for Astrophysics Potsdam (AIP), and the Institute for Astronomy and Astrophysics of the University of Tübingen, with the support of DLR and the Max Planck Society. The Argelander Institute for Astronomy of the University of Bonn and the Ludwig Maximilians Universität Munich also participated in the science preparation for eROSITA.\\
The eROSITA data shown here were processed using the eSASS/NRTA software system developed by the German eROSITA consortium. \\
Funding for the SDSS and SDSS-II has been provided by the Alfred P. Sloan Foundation, the Participating Institutions, the National Science Foundation, the U.S. Department of Energy, the National Aeronautics and Space Administration, the Japanese Monbukagakusho, the Max Planck Society, and the Higher Education Funding Council for England. The SDSS Web Site is http://www.sdss.org/.

The SDSS is managed by the Astrophysical Research Consortium for the Participating Institutions. The Participating Institutions are the American Museum of Natural History, Astrophysical Institute Potsdam, University of Basel, University of Cambridge, Case Western Reserve University, University of Chicago, Drexel University, Fermilab, the Institute for Advanced Study, the Japan Participation Group, Johns Hopkins University, the Joint Institute for Nuclear Astrophysics, the Kavli Institute for Particle Astrophysics and Cosmology, the Korean Scientist Group, the Chinese Academy of Sciences (LAMOST), Los Alamos National Laboratory, the Max-Planck-Institute for Astronomy (MPIA), the Max-Planck-Institute for Astrophysics (MPA), New Mexico State University, Ohio State University, University of Pittsburgh, University of Portsmouth, Princeton University, the United States Naval Observatory, and the University of Washington.
\\

\end{acknowledgements}

\bibliographystyle{aa}
\bibliography{ref.bib}

\begin{appendix} 

\section{AGN models for star-forming and quiescent galaxies}\label{Apd_agn}

The percentages of central star-forming or quiescent galaxies identified as hosting AGN ($f_{\rm AGN}$), based on the BPT diagram, are listed in Table~\ref{Tab_cenagn} \citep{BPT1981,Brinchmann2004}. 
We name the galaxy sample hosting AGN as CEN$_{\rm AGN}$. We follow the method in Paper I to calculate the maximal unresolved AGN emission in the CEN$_{\rm AGN}$ sample ($L_{\rm AGN}$). The AGN luminosity of the CEN or CEN$_{\rm halo}$ is thereby estimated to be $(1-2)\times f_{\rm AGN}\times L_{\rm AGN}$.

\begin{figure}[h!]
\centering
\includegraphics[width=0.9\linewidth]{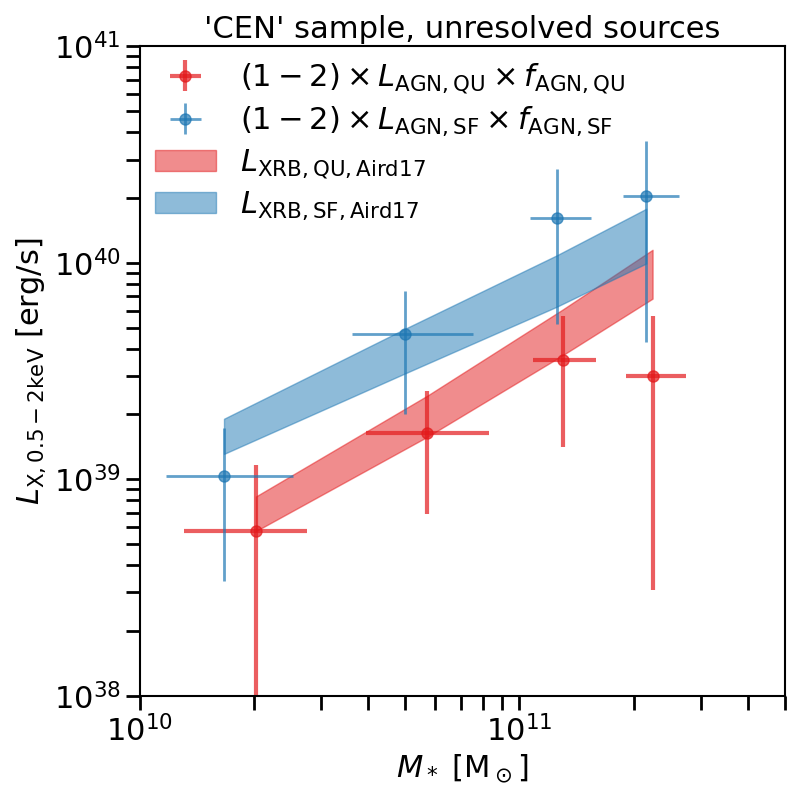}
\includegraphics[width=0.9\linewidth]{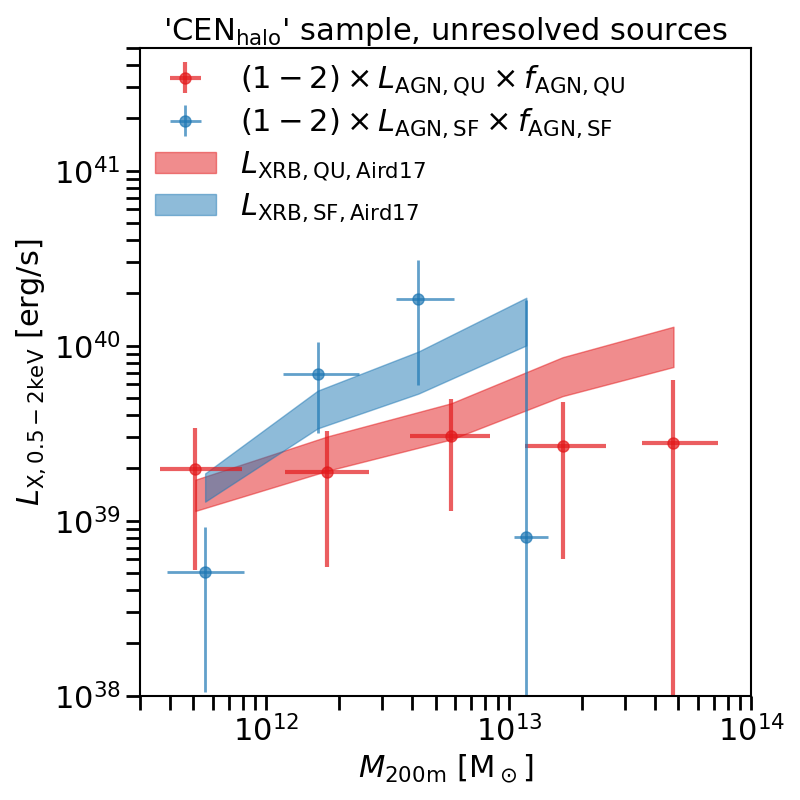}
\caption{Top: the model predicted XRB luminosity (band) and the modeled X-ray luminosity of unresolved AGN (data points) for star-forming (blue) and quiescent (red) galaxies in the CEN sampley. Bottom: Similar to the top panel but for the CEN$_{\rm halo}$ sample.} 
\label{Fig_agn_cen}
\end{figure}

We present in Fig.~\ref{Fig_agn_cen} the modeled X-ray luminosity of XRB and unresolved AGN for CEN and CEN$_{\rm halo}$ samples. 
The star-forming galaxies generally host brighter XRBs and unresolved AGNs than quiescent galaxies.

\begin{table}[h]
\begin{center}
\caption{Percentage and number of galaxies hosting AGN in CEN and CEN$_{\rm halo}$ samples, identified using the BPT diagram \citep{BPT1981,Brinchmann2004}. }
\label{Tab_cenagn}
\begin{tabular}{ccccccccc} 
\hline
\hline
$\log(M_*)$ &$f_{\rm AGN,SF}$ &$N_{\rm CENAGN,SF}$&$f_{\rm AGN,QU}$ &$N_{\rm CENAGN,QU}$\\
\hline
10.0-10.5&13\%&699&22\%&562\\
10.5-11.0&25\%&3377&12\%&2044\\
11.0-11.25&29\%&2367&6\%&1080\\
11.25-11.5&22\%&964&3\%&541\\
\hline
\end{tabular}
\begin{tabular}{ccccccccc} 
$\log(M_{\rm 200m})$ &$f_{\rm AGN,SF}$ &$N_{\rm CENAGN,SF}$&$f_{\rm AGN,QU}$ &$N_{\rm CENAGN,QU}$\\
\hline
11.5-12.0&7\%&1407&19\%&989\\
12.0-12.5&20\%&6081&9\%&1032\\
12.5-13.0&24\%&2880&6\%&1085\\
13.0-13.5&17\%&310&3\%&539\\
13.5-14.0&0\%&0&2\%&170\\
\hline
\end{tabular}
\end{center}
\end{table}

\section{The X-ray surface brightness profiles of star-forming and quiescent galaxies without masking X-ray sources} \label{Apd_nomask}

\begin{figure}
    \centering
    \includegraphics[width=0.9\linewidth]{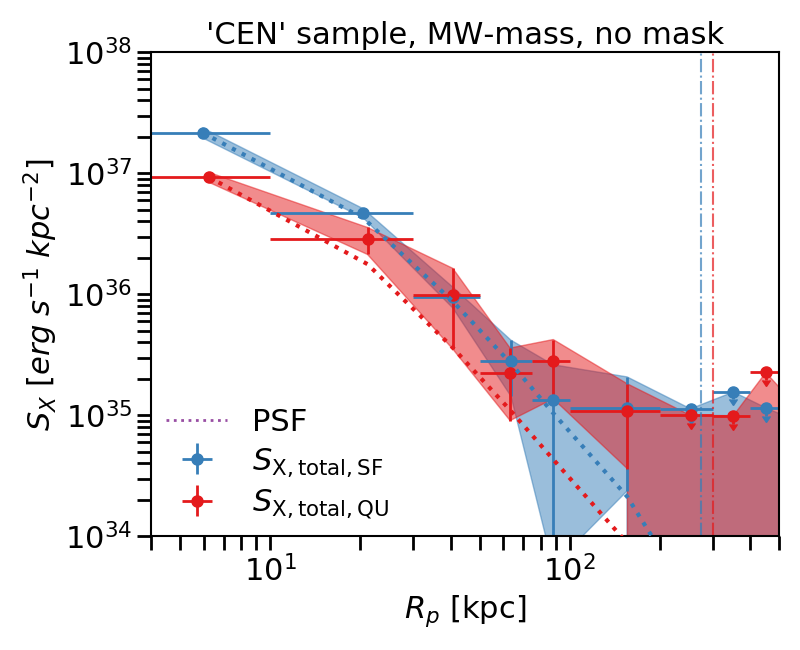}
    \includegraphics[width=0.9\linewidth]{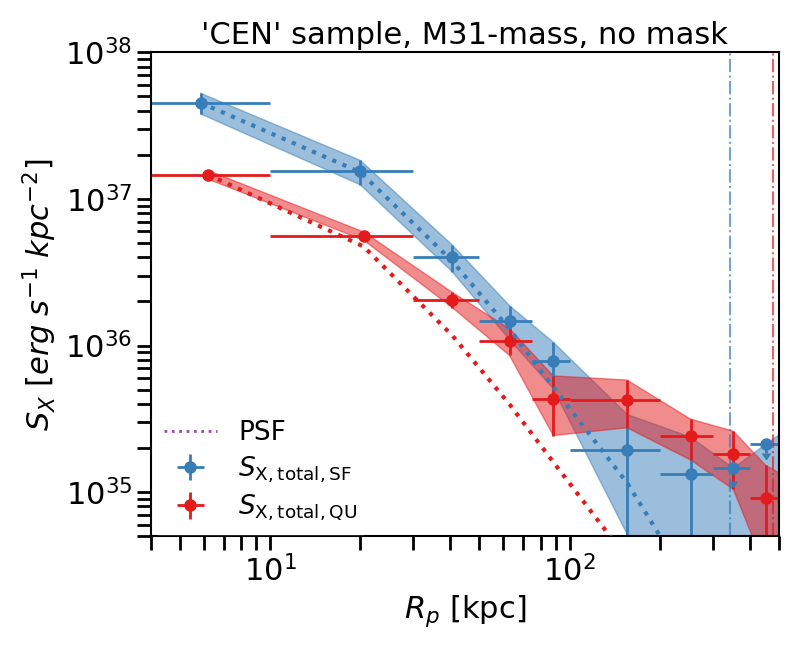} 
    \includegraphics[width=0.9\linewidth]{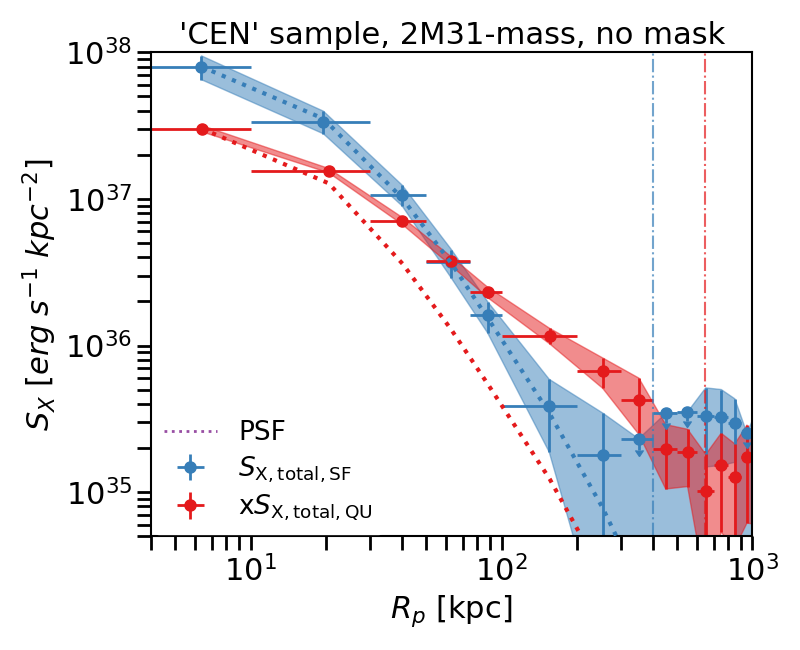}
    \caption{The X-ray surface brightness profiles of central star-forming ($S_{\rm X, mask, SF}$, blue) and quiescent ($S_{\rm X, mask, QU}$, red) galaxies in CEN sample without masking X-ray sources. The vertical dash-dotted line denotes the average virial radius of stacked galaxies.}
        \label{Fig_profile_cen_sfqu_nomask}
\end{figure}

\begin{figure}
    \centering
    \includegraphics[width=0.9\linewidth]{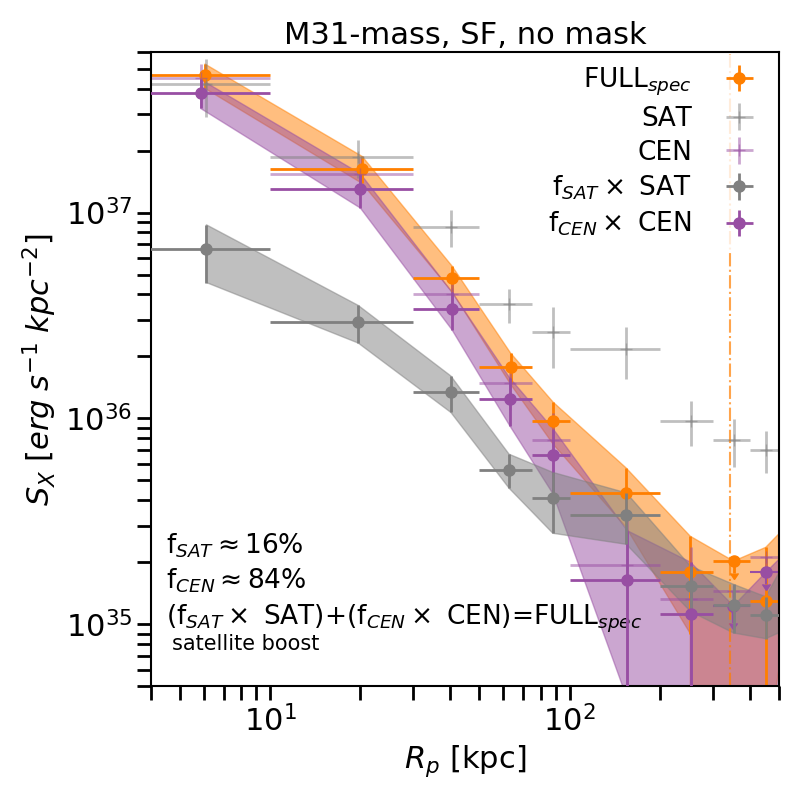}
    \includegraphics[width=0.9\linewidth]{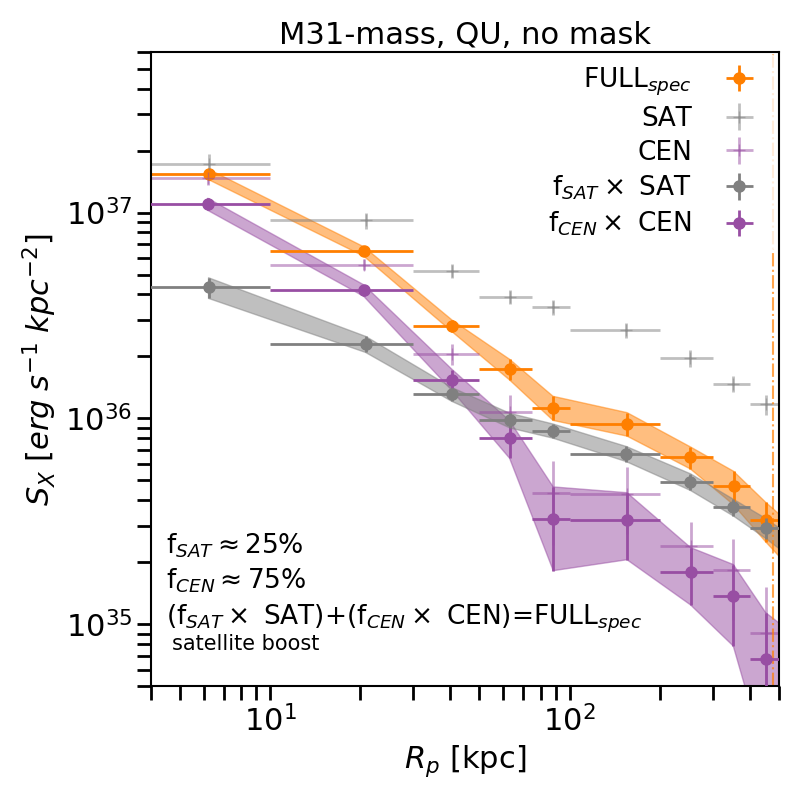} 
    \caption{The X-ray surface brightness profile of galaxies in M31-mass bin of FULL$_{\rm spec}$ sample and the profiles after splitting FULL$_{\rm spec}$ sample to central (CEN) or satellite (SAT) galaxies for star-forming and quiescent galaxies. We scale the observed surface brightness profiles of CEN and SAT according to the number ratio of galaxies, namely, $f_{\rm SAT}=N_{\rm g,SAT}/N_{\rm g,FULL_{spec}}$ and $f_{\rm CEN}=N_{\rm g,CEN}/N_{\rm g,FULL_{spec}}$. The vertical dash-dotted line denotes the average virial radius of M31-mass galaxies in the FULL$_{\rm spec}$ sample. }
        \label{Fig_profile_censat_sfqu_M31}
\end{figure}

We compare the X-ray surface brightness profiles of central star-forming and quiescent galaxies without masking X-ray sources in Fig.~\ref{Fig_profile_cen_sfqu_nomask}. The PSF of eROSITA at the physical scale is overplotted in a dotted line. We find the X-ray surface brightness profile of star-forming galaxies is consistent with the PSF and shows weak excess, while the quiescent galaxies host significant extended X-ray emission. Without introducing the uncertainty of X-ray source masking and unresolved X-ray sources modeling, the more extended X-ray emission around quiescent galaxies supports our result in Sect.~\ref{Sec_cen}.

We discuss in Paper I how the satellite boost biases the central galaxy' X-ray surface brightness profile by stacking the X-ray emission around central (CEN) and satellite (SAT) galaxies and rescaling the profile by the galaxy number fraction, respectively. Here, we repeat the calculation in Paper I for star-forming and quiescent galaxies, respectively. The comparison of the X-ray surface brightness profiles of FULL$_{\rm spec}$ (including CEN and SAT), CEN, and SAT sample and the rescaled profiles are plotted in Fig.~\ref{Fig_profile_censat_sfqu_M31}, taking the M31-mass bin as an example. As expected, galaxies with the same $M_*$ but classified as satellites or centrals have different X-ray surface brightness profiles. The satellite galaxy, residing in a more massive dark matter halo, has its X-ray emission boosted by the emission from virialized gas of the larger halo they reside in, from nearby centrals and other satellite galaxies. Consequently, the satellite galaxies' X-ray surface brightness profile appears to be more extended and brighter than that of the central. 
We find star-forming galaxies suffer less satellite boost bias, as its satellite fraction is low and less extended X-ray emission around it.


\end{appendix}

\end{document}